\documentclass[aps,preprintnumbers, amsmath, showpacs]{revtex4-2}
\usepackage{psfrag}
\usepackage{epsfig}
\usepackage{amsfonts}
\usepackage{amssymb}
\usepackage{graphicx}
\usepackage{subfig}
\usepackage{amsmath}
\usepackage[mathscr]{eucal}

\usepackage[latin1]{inputenc}
\usepackage{color}
\usepackage{suffix}
\usepackage{mathtools}
\usepackage{bm}
\usepackage{soul}

\newcommand{\F}{\mathcal F}
\newcommand{\pd}{\partial}

\newcommand{\bea}{\begin{eqnarray}}
\newcommand{\eea}{\end{eqnarray}}
\newcommand{\beas}{\begin{eqnarray*}}
\newcommand{\eeas}{\end{eqnarray*}}

\def\Title#1{\begin{center} {\Large {\bf #1} } \end{center}}

\begin{document}

\Title{Thermal phonon fluctuations and stability of the magnetic dual chiral density wave phase in dense QCD}

\author{E. J Ferrer, W. Gyory and V. de la Incera}
\affiliation {Dept. of Physics and Astronomy, University of Texas Rio Grande Valley, Edinburg 78539, USA, 
Physics Department, CUNY-Graduate Center, New York 10314, USA}

\begin{abstract}
We study the stability against thermal phonon fluctuations of the magnetic dual chiral density wave (MDCDW) phase, an inhomogeneous phase arising in cold, dense QCD in a magnetic field. Following a recent study that demonstrated the absence of the Landau-Peierls (LP) instability from this phase, we calculate the (threshold) temperature at which the phonon fluctuations wash out the long-range order over a range of magnetic fields and densities relevant to astrophysical applications. Using a high-order Ginzburg-Landau expansion, we find that the threshold temperature is very near the critical temperature for fields of order $10^{18}$ G and still a sizable fraction of the critical temperature for fields of order $10^{17}$ G. Therefore, at sufficiently large magnetic fields, the long-range order of the MDCDW phase is preserved over most of the parameter space where the condensate is energetically favored; at smaller magnetic fields, the long-range order is still maintained over a considerable region of parameter space relevant to compact stars. We provide general symmetry arguments to explain why a magnetic field alone is not enough to eliminate the LP instability that characterizes single-modulated phases in 3+1 dimensions.

\end{abstract}


\maketitle

\section{Introduction}

Mapping the quantum chromodynamics (QCD) phase diagram in the temperature-density plane is currently a significant theoretical and experimental research focus. Thanks to the asymptotic freedom of QCD, perturbative methods can describe the regions of extremely high densities and temperatures, predicting the quark-gluon plasma (QGP) \cite{QGP} phase at high temperatures and relatively low densities and the color superconducting color-flavor locked (CFL) \cite{alf-raj-wil-99/537, CS-Review} phase at low temperatures and high densities. Meanwhile, lattice QCD has become a powerful tool for exploring low-density regions, predicting a ground state with chiral symmetry breaking and color confinement at low temperatures and densities. However, these methods are unavailable for describing the intermediate region of low temperatures and moderately high densities, in the former case due to the large coupling constant, and the latter case due to the sign problem. This region is becoming increasingly important to astrophysics and needs to be better understood, as the conditions within compact stars fall under this part of the phase diagram. To investigate this region, theorists rely on effective theories, nonperturbative methods, and complex simulations. Fortunately, the outcomes of such theoretical and numerical studies can now be compared with a wealth of observational data from multi-messenger astronomy. 

It is worth highlighting that the two physical settings where this region of the QCD phase diagram is relevant- compact stars and heavy-ion collisions (HIC)- typically feature the strongest magnetic fields observed in nature or experiment. The largest surface fields of magnetars are on the order of $ 10^{15}$ G, and inner fields may be several orders larger. Theoretical estimates of the field strength in neutron star (NS) cores based on energy equipartition have yielded upper bounds of order $10^{18}$ G for neutron matter \cite{Nuclear-matter-field} and $10^{20}$ G for quark matter \cite{Quark-Matter-field}. Estimates based on the star stability of slowly rotating NSs give an upper limit of $\sim8\times10^{18}(1.4M_\odot/M)$ G \cite{Cardall2001}. A recent study \cite{Tsokaros} using magnetohydrodynamic simulations in full general relativity applied to rotating NSs found that core field strengths can reach values a few times $10^{17}$ G. Based on these studies, the numerical results of this paper will focus primarily on magnetic field values in the range of $10^{17}$--$10^{18}$ G.

Strong magnetic fields are also produced in off-central HIC experiments, with values reaching $10^{18}-10^{19}$ G immediately after a collision. Currently, no experiments can reach the low-temperature and intermediate-density regions of the phase diagram. However, several upcoming HIC facilities are designed to probe these conditions in the near future. Specifically, the Facility for Antiproton and Ion Research (FAIR) \cite{FAIR} at the GSI site in Germany, and the Nuclotron-based Ion Collider Facility (NICA) \cite{Toneev} at JINR laboratory in Dubna, Russia, will provide new data relevant to cold and dense QCD.

Several studies of quark matter at intermediate densities have found that spatially inhomogeneous condensates are preferred over homogeneous ones \cite{PPNP2015, CSLOFF, CSLOFF-2, Rubakov, Shuster, Park, Rapp, NickelPRL, NickelPRD, Nakano, Kojo, Hidaka, Pisarski, Angel, Kojo-2, CS-MRP86, Gubina, Buballa, Abuki}. For example, QCD in the large-$N_c$ limit \cite{Rubakov}, NJL-like models \cite{NickelPRL, NickelPRD, Nakano}, and quarkyonic matter \cite{Kojo, Hidaka, Pisarski, Kojo-2, Angel} exhibit such phases and find that single-modulated chiral condensates are energetically favored over those with higher-dimensional modulations. However, single-modulated condensates suffer from the so-called Landau-Peierls (LP) instability \cite{LP,HidakaPRD92}. This instability, triggered by the thermal fluctuations in the medium, emerges at any nonzero temperature ($T$), no matter how small. Technically, the LP instability manifests in the existence of a soft mode in the direction transverse to the condensate modulation in the spectrum of the fluctuations. ``Soft" in this context means the spectrum is quadratic rather than linear in the transverse momentum. The softening of the spectrum leads to a divergent average square phonon fluctuation at any finite temperature $T>0.$ As a consequence, the order parameter average vanishes, and the long-range order is washed out. More precisely, a quasi-long-range order remains due to the algebraic decay of the order parameter correlation \cite{HidakaPRD92, Lee-PRD92}, in close analogy with smectic liquid crystals \cite{Smectic}.

Although the LP instability undermines the long-range order of single-modulated inhomogeneous phases in 3+1 dimensions, the situation can, in principle, change in the presence of a magnetic field. Indeed, the magnetic dual chiral density wave (MDCDW) phase is free of LP instability at arbitrarily low temperatures \cite{MDCDW-2}. The key for this to happen, as we will show in this paper, is the coexistence of the rotational symmetry breaking by the external magnetic field with a ground state that breaks time-reversal symmetry. In this ground state, the energy spectrum of the quasiparticles in the lowest Landau level (LLL) is asymmetric about the zero-energy level \cite{KlimenkoPRD82}-\cite{MDCDW-Review}, a feature that leads to nontrivial topological properties  \cite{PLB743}-\cite{MDCDW-Review} and is responsible for the presence of free energy terms that are odd in the condensate modulation $q$. These odd-in-$q$ terms ensure the absence of soft transverse modes in the spectrum of the fluctuations \cite{MDCDW-2}, hence the lack of LP instability. 

The MDCDW phase is energetically preferred over the homogeneous chiral phases (broken and unbroken) in a wide range of the parameter space \cite{MDCDW-3} at supercritical coupling and even at subcritical coupling \cite{Israel}. In sufficiently strong magnetic fields, the condensate is favored at all densities and up to temperatures of tens of MeV, which is several orders of magnitude higher than typical temperatures of old NSs. Moreover, over a range of densities $3.5$--$9$ $n_s$, where $n_s$ is the nuclear saturation density, the condensate magnitude decreases significantly but does not vanish, corresponding to a small ``remnant mass" for the fermion quasiparticles \cite{MDCDW-3}. In addition to these results, the maximum stellar mass of a hybrid star with a core made of quark matter in a variant of the MDCDW phase \cite{InhStars} was shown to be compatible with known stellar mass maximum constraints ($M \gtrsim  2M_{\odot}$) \cite{Demorest, Antoniadis}. Furthermore, at very strong magnetic fields, the speed of sound in this phase reaches values beyond the conformal limit \cite{Aric-2}, consistent with NS expectations \cite{vs}. Finally, the heat capacity of an NS core made of MDCDW matter has been shown \cite{PRD20} to be well above the lower limit expected for NSs ($C_V\gtrsim 10^{36}(T/10^8)$ erg/K) \cite{Cv-NS}. 

The many tests passed by the MDCDW phase so far indicate that it is a robust candidate for the inner phase of NSs with strong internal magnetic fields. However, despite the lack of LP instability,  several important questions on the effects of the thermal fluctuations remain since the latter refers only to the stability at arbitrarily low temperatures. These questions are: (1) At what temperature do the phonon fluctuations become so large that they wash out the long-range order of the MDCDW inhomogeneous phase? We call this temperature the threshold temperature $T_\text{thr}$. (2) Is $T_\text{thr}$ larger than the typical temperatures of old NSs? (3) How does $T_\text{thr}$ compare to the critical temperature $T_c$ at which the chiral condensate evaporates? 

This paper aims to answer the above questions using a very accurate, high-order Ginzburg-Landau (GL) expansion of the MDCDW thermodynamic potential at finite temperature in powers of the order parameter and its derivatives. By exploring the stability of the MDCDW phase at finite temperatures, this study will contribute to the set of essential tests of the viability of this inhomogeneous quark matter phase as a candidate for NS cores. 

The paper is organized as follows. In Sec. \ref{secII}, we develop a method to systematically obtain the low-energy theory of the fluctuations at any desired level of accuracy, taking advantage of the MDCDW GL expansion at arbitrarily large order $N$. Using the low-energy theory, we calculate the average square fluctuation and find the threshold temperature over a range of chemical potentials and magnetic fields. The results are presented and discussed in Sec. \ref{secIII}. In Sec. \ref{secIV}, we give concluding remarks.

\section{Thermal Phonon Fluctuations in the MDCDW Phase}\label{secII}
 In this section, we develop a systematic method to compute the average of the thermal fluctuations at any desired accuracy. It is based on a known procedure that starts with the MDCDW generalized GL expansion, then considers the ground state perturbed by low-energy fluctuations, and finally expands up to quadratic order in the fluctuation to find the low-energy theory of the fluctuations. Using this low-energy theory of the fluctuations, one can readily calculate the average square fluctuation. From the derivations in \cite{MDCDW-3}, we know that physical parameters found using the GL expansion of order 20 and higher overlap very accurately with the exact numerical calculations and, hence, are physically reliable.

For clarity, we will start by reviewing the procedure using the sixth-order GL expansion, as laid out in \cite{MDCDW-2}, and then generalize the approach to an arbitrarily order $N$. Since all of the coefficients of the GL expansion can be quickly computed using the formulas derived in \cite{MDCDW-3}, our approach will provide a fast and systematic way to derive the low-energy theory of the fluctuations with any desired accuracy. We will then use these formulas to calculate the average square fluctuation and the threshold temperature. 

\subsection{Low-energy theory of fluctuations from the sixth-order GL expansion}
\label{secIIA}
 
Let us consider a two-flavor effective theory of interacting quarks described by the following NJL model in an external magnetic field, 
\begin{eqnarray} \label{L_NJL_QED}
\mathcal{L}=\bar{\psi}[i\gamma^{\mu}(\partial_\mu+iQA_{\mu})]\psi 
+G[(\bar{\psi}\psi)^2+(\bar{\psi}i\tau\gamma_5\psi)^2].
\end{eqnarray}
Here, $Q=\mathrm{diag} (e_u,e_d)=\mathrm{diag} (\frac{2}{3}e,-\frac{1}{3}e)$, $\psi^T=(u,d)$, $\tau=(\tau_1,\tau_2,\tau_3)$ are Pauli matrices in flavor space, and $G$ is the four-fermion coupling. The electromagnetic potential $A^{\mu}=(0,0,Bx,0)$ corresponds to a constant and uniform magnetic field $\mathbf{B}$ pointing in the $z$-direction, with $x^{\mu}=(t,x,y,z)$.  The presence of $B$ explicitly breaks the isospin symmetry and the rotational symmetry about the $x$ and $y$ axes. Hence, the global continuous symmetry of the theory is $U(1)_V\times U(1)_A \times SO(2)\times R^3$.

One can now investigate the ground state of this theory at finite baryon density using a generalized GL expansion in powers of a single-modulated order parameter $M(z)=\sigma +i \pi$ and its derivatives, with $\sigma= -2G\langle\bar{\psi}\psi\rangle$ and $\pi=-2G\langle\bar{\psi} i\gamma^5\tau_3\psi\rangle$ scalar and pseudoscalar condensates respectively. The sixth-order GL expansion takes the form
\begin{eqnarray} \label{GL1}
\Omega^{(6)}&=& a_{2,0}|M|^{2}-i \frac{b_{3,1}}{2}[M^{*}(\hat{B} \cdot \nabla M)-(\hat{B} \cdot \nabla M^{*}) M]+a_{4,0}|M|^{4}+a_{4,2}|\nabla M|^{2} \nonumber 
\\ 
&-&i\frac{b_{5,1}}{2}|M|^{2}[M^{*}(\hat{B} \cdot \nabla M)-(\hat{B} \cdot \nabla M^{*}) M]+\frac{i b_{5,3}}{2}[(\nabla^{2} M^{*}) \hat{B} \cdot \nabla M-\hat{B} \cdot \nabla M^{*}(\nabla^{2} M)] 
\nonumber
\\
&+&a_{6,0}|M|^{6}+a_{6,2}|M|^{2}|\nabla M|^{2}+a_{6,4}|\nabla^{2} M|^{2},
\end{eqnarray}
where $\hat{B}=B/|B|$, and the coefficients $a_{i,j}$ and $b_{i,j}$ are functions of the temperature, quark chemical potential $\mu$, and magnetic field. The GL expansion should respect all the global symmetries of the original theory, including the discrete ones.

Notice that the structures with $b$ coefficients are odd in the gradient of the order parameter and, therefore, odd in the condensate's modulation. They can only exist in the presence of an external field, that is, when the isotropy of the original theory is explicitly broken, even though, as it will become clear below, the explicit breaking of the isotropy is a necessary but not sufficient condition. In this situation, the rotational symmetry reduces to rotations about an axis parallel to the external field. Thus, all the $b_{i,j}$ should vanish when $B\to 0$. Indeed, as confirmed by explicit calculations \cite{MDCDW-3}, in the MDCDW theory, the $b$ coefficients vanish in the limit $B\to 0$. Moreover, since $B$ is odd under time ($T$) reversal symmetry, the order parameter should satisfy $M \leftrightarrow M^{*}$ under $T$ to ensure the invariance of these terms under $T$. This is possible here since the order parameter's imaginary part $\pi$ is odd under $T$. 

The MDCDW condensate ansatz is
\begin{equation}
\label{ansatz}
M(z)=-2G\Delta e^{iqz}=me^{iqz}.
\end{equation} 
It spontaneously breaks chiral and translational symmetries. Notice that the $T$ transformation of the condensate is equivalent to changing $q \to -q$. Accordingly, once we use this ansatz in (\ref{GL1}), the $T$-invariance of the resulting expression will turn out to be invariance under $q \to -q$ together with $B \to -B$.

Notice that one does not need to include additional $B$-dependent structures that are even in $q$, like those with coefficients $a_{4,2}^{(1)}$ and $a_{6,2}^{(1)}$ in \cite{MDCDW-2}, as those structures would be redundant in this ansatz. The reason is that the modulation of the MDCDW condensate is parallel to the magnetic field; hence, such $B$-dependent structures will not lead to new linearly independent terms.

Since the structures with $b$ coefficients are odd in the condensate modulation $q$, they must originate from the LLL part of the thermodynamic potential. To see this, consider the spectrum of the fermion quasiparticles in the MDCDW theory,
\begin{align}
    E^0 &= \epsilon\sqrt{m^2+k^2}+q/2 & \epsilon&=\pm1,\quad \ell=0
    \label{LLL}
    \\
    E^\ell &= \epsilon\sqrt{\Big(\xi\sqrt{m^2+k^2}+q/2\Big)^2+2|e_fB|\ell} & \epsilon,\xi&=\pm1,\quad \ell=1,2,3,\dots.
    \label{HLL}
\end{align}
The higher Landau level (HLL) modes, given by (\ref{HLL}) with $\xi=\pm$ the spin projection and $\epsilon$ indicating positive/negative energies, are symmetric about $E=0$. For the HLL, changing $q \to -q$ exchanges the HLL modes, leaving the part of the thermodynamic potential that depends on the HLL spectrum invariant. Therefore, the HLL cannot contribute to the $b$ terms. In contrast, the LLL spectrum is asymmetric about $E=0$ \cite{KlimenkoPRD82}, and $\epsilon$ does not indicate positive or negative energy. For example, for $m<q/2$, both modes can be nonnegative. Under $q \to -q$, the LLL modes do not transform into each other, and so the part of the thermodynamic potential that depends on these modes is not invariant under such a transformation; thus, the $b$-structures in (\ref{GL1}) can only come from the LLL spectrum \cite{MDCDW-3}. 

Most theories of fermions in a magnetic field have a symmetric spectrum for all the Landau levels, so one might wonder what makes the MDCDW phase different. The answer lies in the \textit{coexistence} of a ground state that breaks $T$ and the dimensional reduction of the LLL in a magnetic field. Because of these two factors, the LLL dynamics in the MDCDW phase resemble the fermion dynamics of the so-called chiral spiral crystal phase (CSCP) in NJL$_2$, a renormalizable model in 1+1 dimensions \cite{Basaretal}. The chiral spiral crystal condensate is the energetically favored solution of that system. This condensate breaks $T$ and gives rise to an asymmetric fermion spectrum of the same form as (\ref{LLL}). Even more strikingly, a close look at the GL expansion of the NJL$_2$ theory in the CSCP \cite{Basaretal} shows that the coefficients that multiply the terms odd in the modulation, denoted by $a_{n_b+2}$ with $n_b$ odd \cite{Basaretal}, are proportional to the same functions of $\mu$ and $T$ as the $\beta_{n_b+2,n_b}$ coefficients in the MDCDW case \cite{MDCDW-3}. The $\beta$ coefficients in \cite{MDCDW-3} correspond to the $b$ coefficients in the notation used in (\ref{GL1}), up to factors of $2^{n_b}$ due to expanding in $q$ instead of in $b=q/2$. The asymmetric spectrum of MDCDW is responsible for the nontrivial topology that manifests in the existence of the $b$-terms in the GL expansion and in several topological properties, such as an anomalous baryon number, anomalous Hall conductivity, and generation of axion polaritons due to the interaction of electromagnetic waves with the MDCDW medium  \cite{PLB743}-\cite{ MDCDW-Review}. 

We call attention to the fact that without a $T$-breaking ground state, no $b$-terms would be present, and since these terms are a direct consequence of the LLL spectral asymmetry, their absence means that the LLL spectrum would have to be symmetric in this case, nowithstanding the dimensional reduction. This can also be gathered from the analogy with the real kink crystal (RKC) in the NJL$_2$ case. The RKC is a $T$-even spatially inhomogeneous condensate that solves the NJL$_2$ equations, although it is not the most energetically favored \cite{Basaretal}. Its GL expansion has no terms odd in the condensate derivative, i.e., it has no b-terms. The spectrum of the fermion quasiparticles in the RKC is symmetric. This example illustrates that the reduced dimension of the fermion dynamics alone is insufficient to ensure an asymmetric spectrum. 

In terms of $m$ and $q$, the sixth-order GL expansion (\ref{GL1}) becomes
\begin{align}
\label{GL2}
\Omega^{(6)}=&a_{2,0}m^2+b_{3,1}m^2q+a_{4,0}m^4+a_{4,2}m^2q^2+b_{5,1}m^4q\nonumber\\
&+b_{5,3}m^2q^3+a_{6,0}m^6+a_{6,2} m^4q^2+a_{6,4}m^2q^4.
\end{align}
The coefficients $a_{i,j}$ and $b_{i,j}$ are found from the MDCDW thermodynamic potential. Explicit expressions in terms of the magnetic field, chemical potential, and temperature were found in \cite{MDCDW-3}. 

To find the ground state solution, we minimize (\ref{GL2}) with respect to the dynamical parameters 
\begin{subequations}
\begin{align}
\partial \Omega^{(6)}/\partial m&=&2m \{ a_{2,0}+2a_{4,0} m^2+3a_{6,0}m^4+q^2[a_{4,2}+2a_{6,2}m^2+a_{6,4}q^2]
+q [b_{3,1}+2b_{5,1} m^2+b_{5,3}q^2]\}=0
\label{SC-Delta}
\\
\partial \Omega^{(6)}/\partial q&=&2qm^2 \{a_{4.2}+a_{6,2}m^2+2a_{6.4}q^2 +\frac{b_{3,1}}{2q}+b_{5,1} \frac{m^2}{2q}+\frac{3}{2}b_{5,3}q\}=0 \qquad \qquad \qquad\qquad \qquad\quad \qquad
\label{SC-q}
\end{align}
\end{subequations}
 Using the expressions for the $a$ and $b$ coefficients calculated in \cite{MDCDW-3}, we can numerically find the minimum solutions $m_0$ and $q_0$ as functions of the magnetic field, temperature, and chemical potential.

Next, we derive the low-energy theory of the fluctuations. When $m_0$ and $q_0$ are nonzero, the chiral and translational symmetries are spontaneously broken, giving rise to two Nambu-Goldstone (NG) bosons: a neutral pion and a phonon. Under the effect of these NG bosons, the MDCDW condensate changes as
\begin{equation}\label{SB-Pattern-Finite}
M(x) \rightarrow e^{i\tau}M(z+\xi)=e^{i(\tau+q\xi)}M(z)
 \end{equation}
It is easy to see that these transformations are locked, as discussed in \cite{MDCDW-2} and references therein. Consequently, in this case, only one independent NG boson, a mix of pions and phonons, exists. Thus, although one can formally define the fluctuation of the condensate's phase as a phonon, a pion, or a combination of them, this is just a formality, as due to the mixing, one cannot distinguish among them. For convenience, and following \cite{MDCDW-2}, we use the phonon here.

The low-energy theory of fluctuations can now be found by first considering the ground state solution $M_0(z)=m_0e^{iq_0z}$ with its phase perturbed by a small phonon fluctuation $u=u(x,y,z)$ and expanded in powers of $u$,
\begin{align}
\label{perturbation}
M(x) &= M[z+u(x,y,z)]\nonumber\\
&\approx M_0(z)+M_0'(z)u+\frac12M_0''(z)u^2+\cdots.
\end{align}

We then insert (\ref{perturbation}) into (\ref{GL1}) and expand the resulting expression in powers of $u(x,y,z)$ up to the second order. Since we are interested in the infrared region, we can neglect higher-order terms in momentum. The result of this process is the free energy density of the fluctuations 
\begin{equation}
\label{F(6)}
  \F^{(6)}[u(x,y,z)]= \F^{(6)}_0+m^2q^2v_z^2(\pd_z u)^2+m^2q^2v_\perp^2(\pd_\perp u)^2+\mathcal{O}((\nabla^2 u)^2)
  \end{equation}
where $(\pd_\perp u)^2=(\pd_x u)^2+(\pd_y u)^2$, $\F^{(6)}_0=\Omega^{(6)}(M_0)$, and for simplicity, we used $m$ and $q$ to denote the minimum solutions $m_0$ and $q_0$ and shall keep that notation from now on. In deriving (\ref{F(6)}), a term linear in $\pd_z u$ appears, but its coefficient is proportional to the stationary equation (\ref{SC-q}) and hence vanishes. The group velocities $v_z$ and $v_\perp$ are functions of the GL coefficients and powers of the ground state solutions $m$ and $q$. Their specific expressions will depend on the order of the GL expansion used. In the sixth-order case 
\begin{align}
    v_z^2&=a_{4,2}+m^2a_{6,2}+6q^2a_{6,4}+3qb_{5,3} \label{vz(6)}\\
    v_\perp^2&=a_{4,2}+m^2a_{6,2}+2q^2a_{6,4} +qb_{5,3}\label{vperp(6)}.
\end{align}

Notice that $v_\perp^2$ can be written as
\begin{equation}
v_\perp^2= \frac{1}{2qm^2} \frac{\partial \Omega^{(6)}}{\partial q} -\frac{b_{3,1}}{2q}-b_{5,1} \frac{m^2}{2q}-\frac{1}{2}qb_{5,3}=-\frac{b_{3,1}}{2q}-b_{5,1} \frac{m^2}{2q}-\frac{1}{2}qb_{5,3}
\label{vperp-v2},
\end{equation}
thanks to the stationary condition (\ref{SC-q}). Written in this form, it is evident that the presence of the $b$ coefficients is essential for a non-vanishing transverse group velocity. On the other hand, it is easy to see that neither of the stationary conditions (\ref{SC-Delta})-(\ref{SC-q}) can make $v_z^2$ equal zero. The fact that $v_z^2$ is nonzero and depends on the $a$ and $b$ coefficients indicates that it gets contributions from all the LLs. Explicit profiles of the velocities vs. the chemical potential at various magnetic fields will be presented in Fig. \ref{velocities} of Section \ref{secIII}.

It is now convenient to rewrite the fluctuation field $u$ in terms of the axion field $\theta \equiv \theta(x,y,z)=mqu$ and express (\ref{F(6)}) in terms of $\theta$. The corresponding Lagrangian density of the low-energy effective theory of the fluctuations can then be written as

\begin{equation}
\label{L(6)}
\mathcal{L}^{(6)}=\frac{1}{2}[(\pd_0 \theta)^2-v_z^2(\pd_z \theta)^2-v_\perp^2(\pd_\perp \theta)^2-\xi^2(\nabla^2 \theta)^2]
\end{equation}
Accordingly, the spectrum of the fluctuation field is found to be anisotropic and linear in the longitudinal and transverse directions
\begin{equation}\label{theta spectrum}
E\simeq\sqrt{v^2_zk^2_z +v_\bot^2k_\bot^2},
\end{equation}
where $k_\bot^2=k_x^2+k_y^2$.

To explore the effects of the fluctuations on the condensate, one considers its average
\begin{equation}\label{averageM}
\langle M \rangle=m e^{iqz}\langle \cos qu \rangle,
\end{equation}
where 
 \begin{equation}\label{average}
\langle  ... \rangle =\frac{\int \mathcal{D}u(x) ... e^{-S(u^2)}}{\int \mathcal{D}u(x) e^{-S(u^2)}}
\end{equation}
with
 \begin{equation}\label{phonon action}
S(u^2)= T\sum_n \int^{\infty}_{-\infty} \frac{d^3k}{(2\pi)^3} [\omega^2_n+(v^2_zk^2_z +v_\bot^2k_\bot^2)]q^2m^2u^2
\end{equation}
the finite-temperature, low-energy effective action of the phonon and $\omega_n=2n\pi T$ the Matsubara frequency. 

We can now consider the relation
\begin{equation}\label{cos-exp-relation}
\langle \cos qu \rangle =e^{-\langle q^2u^2 \rangle/2}
\end{equation}
and use (\ref{average}) to find the mean square of the fluctuation in the low-energy region as
\begin{eqnarray}\label{<q2u2>(6)}
\langle q^2u^2 \rangle &=&\frac{1}{(2\pi)^2}\int_{0}^\infty dk_\bot k_\bot \int_{-\infty}^\infty dk_z \frac{T}{m^2(v^2_zk^2_z +v_\bot^2k_\bot^2)}\nonumber
\\
&\simeq& \frac{T}{8\pi m\sqrt{v^2_zv_\bot^2}},
\end{eqnarray}
Here, we considered that the main contribution comes from the infrared region, where the lowest Matsubara mode is dominant. More details of this calculation can be found in \cite{MDCDW-2}.  A misprint in the mean square formula in \cite{MDCDW-2} has been corrected here.

The result  (\ref{<q2u2>(6)}) is meaningful as long as the inhomogeneous condensate exists because these are phonon fluctuations of the inhomogeneous condensate. Likewise, this expression is only valid for temperatures smaller than the critical temperature $T_c(\mu)$ at which the condensate melts and the system transitions to the chirally restored phase.  Assuming these conditions are met, as long as $\sqrt{v^2_zv_\bot^2} \neq 0$, the fluctuation is finite, and the condensate average does not vanish. From Eq. (\ref{vperp-v2}) and the paragraph below, one sees that this is indeed the case, at least for sufficiently small $T$. We conclude that thanks to the $b$-coefficients, there is no LP instability in the MDCDW phase, and thus, the long-range order of this phase is maintained at least at low temperatures \cite{MDCDW-2}. 

Given that the $b$-coefficients arise from the asymmetric spectrum of the LLL, which is directly related to the nontrivial topology of the LLL fermion dynamics, it is reasonable to deduce that the lack of the LP instability in the MDCDW phase is a consequence of that same topology; in other words, this topology makes the condensate and hence the long-range order robust in the presence of fluctuations. Considering that the $b$-terms result from the coexistence of the dimensional reduction of the LLL and a ground state that breaks the $T$-symmetry, one can infer that a magnetic field alone is not enough to avoid the LP instability. In consequence, we expect that, for instance, a magnetic field would not eliminate the LP instability of the solitonic phase \cite{NickelPRL, PPNP2015} characterized by a $T$-even, single-modulated real condensate in 3+1-dimensions.

On the other hand, let us see what happens when the ground state breaks the $T$ symmetry, but there is no magnetic field. For that, let us consider the DCDW phase, whose ground state is a density wave condensate of the same form as in the MDCDW, only without a magnetic field. The DCDW GL expansion has the same form as Eq. (\ref{GL2}), except there are no $b$-terms because the fermion spectrum in this phase is symmetric about zero. If $B=0$, the expression for $v^2_z$ in (\ref{vz(6)}) becomes a combination of $a$ coefficients that is still nonvanishing at low $T$. However, the situation is different for the transverse velocity, which reduces to 
\begin{equation}\label{B=0(6)}
v_\perp^2=a_{4,2}+m^2a_{6,2}+2q^2a_{6,4}=\frac{1}{2m^2q}\frac{\partial \Omega^{(6)}}{\partial q}=0.
\end{equation}
In this case, the spectrum of the fluctuations softens in the transverse direction, i.e., it is no longer linear along that direction. As a consequence, the denominator in (\ref{<q2u2>(6)}) is zero, the average square fluctuation diverges, and the average of the condensate vanishes. This is the essence of the LP instability that affects all the single-modulated phases in 3+1 dimensions at $B=0$. This example underlines the relevance of the coexistence of the two features already mentioned to ensure the lack of LP instability.  At finite temperature, the long-range order of the DCDW phase is replaced by a quasi-long-range order characterized by algebraically decaying correlation functions at large distances \cite{HidakaPRD92, Lee-PRD92}. 

Although it is clear that there is no LP instability in the MDCDW phase, an important question remains: At what temperature do the phonon fluctuations become so large that they effectively erase the long-range order of the MDCDW inhomogeneous phase? As mentioned in the Introduction, we call this temperature the threshold temperature $T_\text{thr}$ and define it as the temperature where $\langle M \rangle=e^{-1}M_0$. Hence, taking into account (\ref{averageM}), (\ref{cos-exp-relation}) and (\ref{<q2u2>(6)}) it follows that
\begin{equation}\label{T_thr-demo}
T_\text{thr} =16\pi m |v_z| |v_\bot|.
\end{equation} 
Eq. (\ref{T_thr-demo}) is a transcendental equation since $m$, $v_z$, and $v_\perp$ depend on temperature. Thus, it must be solved numerically.

\subsection{Low-energy theory of fluctuations from the Nth-order GL expansion}
\label{SecIIB}
 
The results outlined in the preceding section are limited by the accuracy of the sixth-order GL expansion. As shown in \cite{MDCDW-3}, the sixth-order expansion does not reproduce the correct values of the physical parameters found from the exact thermodynamic potential. We must consider higher-order terms in the GL expansion to obtain reliable results. To understand why the sixth-order expansion is not very accurate, let us recall that the generalized GL expansion is an expansion in powers of the condensate magnitude and its derivatives. While the magnitude is always much smaller than all the parameters near the transition line to the chirally symmetric region, the condensate's derivative is not. For the MDCDW condensate, the expansion in powers of the derivatives becomes an expansion in powers of $q/2$, with $q/2 \lesssim \mu$. However, such an expansion turns out to be an expansion in powers of $\frac{q}{2\mu}$ after writing out the explicit expressions for the coefficients. Hence, to obtain reliable results in the inhomogeneous phase, one must go to a much higher order in the GL expansion than one would normally do for homogeneous ground states. Fortunately, the physical results found using a GL expansion of order 20 and higher are indistinguishable from those found with an exact numerical calculation \cite{MDCDW-3}. This indicates that, in order to find reliable results for the effect of the fluctuations, we must first extend the method of the previous section, considering a GL expansion at arbitrary order $N$.

The step in the previous section in which (\ref{perturbation}) is inserted into (\ref{GL1}) becomes tedious at higher GL expansion orders since there are many more terms in (\ref{GL1}), and these terms contain more derivatives. In this section, we overcome this issue by generalizing the preceding calculation to the case of an arbitrary $N$th-order GL expansion. First, we observe from (\ref{GL1}) that each term in the GL expansion can be written as one of the following expressions,
\begin{align}
\label{a terms}
    &a_{n,n_q}|M|^{n-2-n_q}|\nabla^{n_q/2}M|^2,
    &n=2,4,6,\dots,\quad n_q=0,2,4,\dots,n-2 
    \\ \label{b terms}
    &b_{n,n_q}|M|^{n-2-n_q}i^{n_q-1}\operatorname{Im}[(\nabla^{n_q-1}M^*)(\hat z\cdot\nabla M)]
    &n=3,5,7,\dots,\quad n_q=1,3,5,\dots,n-2,
\end{align}
depending on whether $n$ is even or odd, respectively. Here, we have used $\hat z$ instead of $\hat B$ to denote the unit vector in the $z$-direction. 

We are only considering the low-energy theory, so powers of momentum $k^n$ for $n\geqslant3$ can be neglected, as the expansion is dominated by $k^2$ in the infrared.  Therefore, in calculating $\nabla^kM$, we must only keep track of terms with at most two overall derivatives of $u$, i.e., terms of the form $\pd_iu$, $\pd_i\pd_ju,$ and $(\pd_iu)(\pd_ju)$. To see this in the context of the sixth-order calculation, note that in deriving (\ref{F(6)}), we neglected a term $a_{6,4}m^2q^2(\pd_z^2u+\pd_\perp^2u)^2$, since this would only change the denominator in the integrand of (\ref{<q2u2>(6)}) to $v_z^2k_z^2+v_\perp^2k_\perp^2+a_{6,4}k^4$, and the additional $k^4$ term is negligible in the infrared.


By applying derivatives to $M$ repeatedly, keeping only the relevant terms discussed above, and using an induction argument, one can show that
\begin{align}
\label{d^kM2}
    |\nabla^kM|^2&=m^2 q^{2k}|\hat z+\nabla u|^{2k}, & k&=0,1,2,\dots \\
    \label{Im d^kM}
    \operatorname{Im}[(\nabla^k M^*)(\hat z\cdot\nabla M)]&=m^2 q^{k+1}(-i)^k |\hat z+\nabla u|^k(1+\pd_zu) , & k&=0,2,4,\dots.
\end{align}
The details of this calculation are given in Appendix \ref{app:d^k M}.

The expressions in (\ref{d^kM2}) and (\ref{Im d^kM}) can be further simplified by expanding $|\hat z+\nabla u|^k$, where $k$ is even, and keeping only terms up to second order in $u$, 
\begin{align}
\label{(z+du)^k}
    |\hat z+\nabla u|^k
    &=(1+2\pd_zu+|\nabla u|^2)^{k/2}\nonumber\\
    &=1+k\pd_zu+\frac k2|\nabla u|^2+k\left(\frac k2-1\right)(\pd_zu)^2.
\end{align}
%


Inserting (\ref{(z+du)^k}) into (\ref{d^kM2}) and (\ref{Im d^kM}), again keeping only the relevant terms, and applying the resulting formulas to (\ref{a terms}) and (\ref{b terms}), it is now straightforward to show that the free-energy density of the fluctuations can be written as
\begin{equation} \label{Fn}
    \F^{(N)}=\sum_{n,n_q}^N c_{n,n_q}m^{n-n_q}q^{n_q}\left[1+n_q\pd_zu+\frac{n_q(n_q-1)}2(\pd_zu)^2+\left\lfloor \frac{n_q}2\right\rfloor(\pd_\perp u)^2\right],
\end{equation}
where $c_{n,n_q}$ denotes $a_{n,n_q}$ or $b_{n,n_q}$ if $n$ is even or odd, respectively, and $\lfloor\cdot\rfloor$ is the floor function. The sum includes all pairs $(n,n_q)$ indicated in (\ref{a terms}) and (\ref{b terms}) with $n\leq N$.

Distributing the sum, the first term is just the mean-field GL expansion evaluated at the minimum solutions $m$ and $q$, which we denote $\F_0^{(N)}$. The next term is linear in $\pd_zu$, with a factor of $\sum c_{n,n_q}m^{n-n_q}q^{n_q}n_q$. However, the stationary equation in $q$ gives 
\begin{equation}
\label{stationary(N)}
    \pd\Omega^{(N)}/\pd q=\sum_{n,n_q}^N c_{n,n_q}m^{n-n_q}q^{n_q-1}n_q=0.
\end{equation}
Multiplying (\ref{stationary(N)}) by $q$, we see that the factor multiplying $\pd_zu$  in (\ref {Fn}) vanishes. Rewriting the fluctuation $u$ in terms of the axion field $\theta=mqu$, we have
\begin{equation}
    \F^{(N)}=\F_0^{(N)}+v_z^2(\pd_z\theta)^2+v_\perp^2(\pd_\perp\theta)^2,
\end{equation}
where
\begin{align}
\label{vz(N)}
    v_z^2&=\sum_{n,n_q}^N c_{n,n_q}m^{n-n_q-2}q^{n_q-2}\frac{n_q(n_q-1)}2\\
    \label{vperp(N)}
    v_\perp^2&=\sum_{n,n_q}^N c_{n,n_q}m^{n-n_q-2}q^{n_q-2}\left\lfloor\frac{n_q}2\right\rfloor.
\end{align}
One can easily check that in the case $N=6$, (\ref{vz(N)}) and (\ref{vperp(N)}) reproduce the sixth-order expressions for $v_z$ and $v_\perp$ found in (\ref{vz(6)}) and (\ref{vperp(6)}).

Finally, we verify that $v_\perp$ vanishes in the $B\to 0$ limit. When $B=0$, the $b$ coefficients vanish, so the only nonzero terms in (\ref{vperp(N)}) are those for which $n_q$ is even. We can then rewrite $\lfloor n_q/2\rfloor$ as $n_q/2$. Comparing with (\ref{stationary(N)}), we see that 
\begin{equation}\label{vperp(N)B=0}
v_\perp^2=\frac{1}{2m^2q}\frac{\partial \Omega^{(N)}}{\partial q}=0,
\end{equation}
in agreement with the sixth-order case shown in (\ref{B=0(6)}). Again, the LP instability returns in the $B\to 0$ limit for the inhomogeneous phase.

The threshold temperature formula, Eq.(\ref{T_thr-demo}), is valid at any order $N$ of the GL expansion. However, the $T_\text{thr}$ solution depends on the GL expansion coefficients and thus on the expansion order through the group velocities and $m$.  We know the expansion is accurate once all the curves found after a certain order $N$ overlap entirely in the relevant density region. As we will see in the next section, that happens at order 20th and higher, a result consistent with the exact numerical calculations performed in \cite{MDCDW-3}.

\section{Numerical Results}
\label{secIII}
 

To answer the questions in the Introduction, we start by finding the profile of the threshold temperature vs. chemical potential calculated using GL expansions at different orders. This lets us identify the approximation where the curves entirely overlap and hence can be trusted. The formulas found in Sec. \ref{SecIIB}, combined with the explicit formulas for the GL coefficients in \cite{MDCDW-3}, enable fast numerical computation of the threshold temperature at any arbitrary order $N$. As in \cite{MDCDW-3}, we work in the chiral limit and use proper-time regularization with $\Lambda\approx636.790$ MeV and coupling constant $G\Lambda^2=6$; these parameter values correspond to $m_\text{vac}=300$ MeV.

\begin{figure}[h]
\includegraphics[width=.42\textwidth]{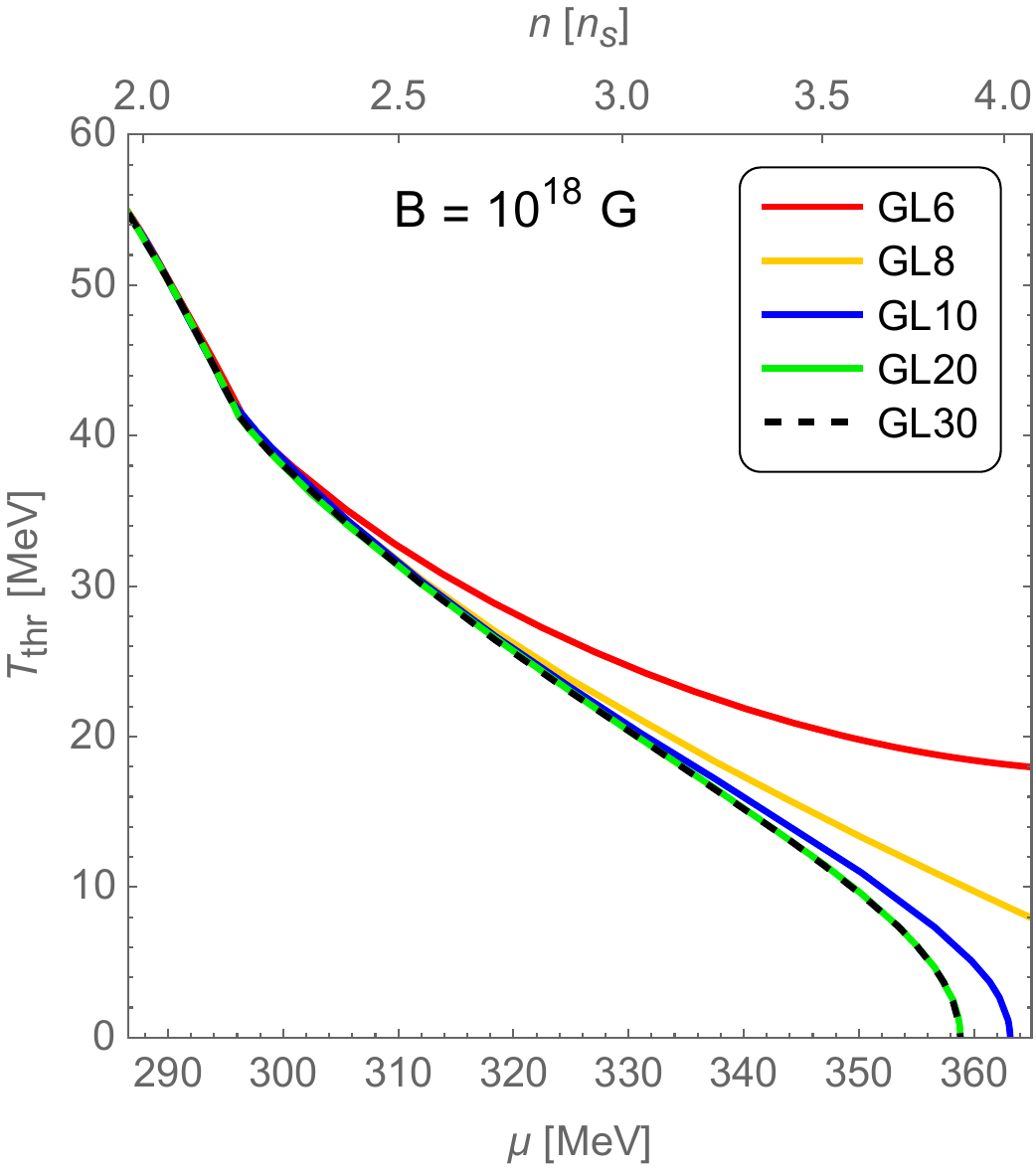}\hfill
 \caption{Threshold temperature at $B=10^{18}$ G vs. quark chemical potential (bottom axis) and approximate density (top axis), computed using the $N$th-order GL expansion for $N=6,8,10,20,30$. Note that we need an order 20th or higher to obtain accurate results}
  \label{GL6to30}
\end{figure}
 
 \begin{figure}
\centering
    \subfloat{
    \includegraphics[width=.42\textwidth]{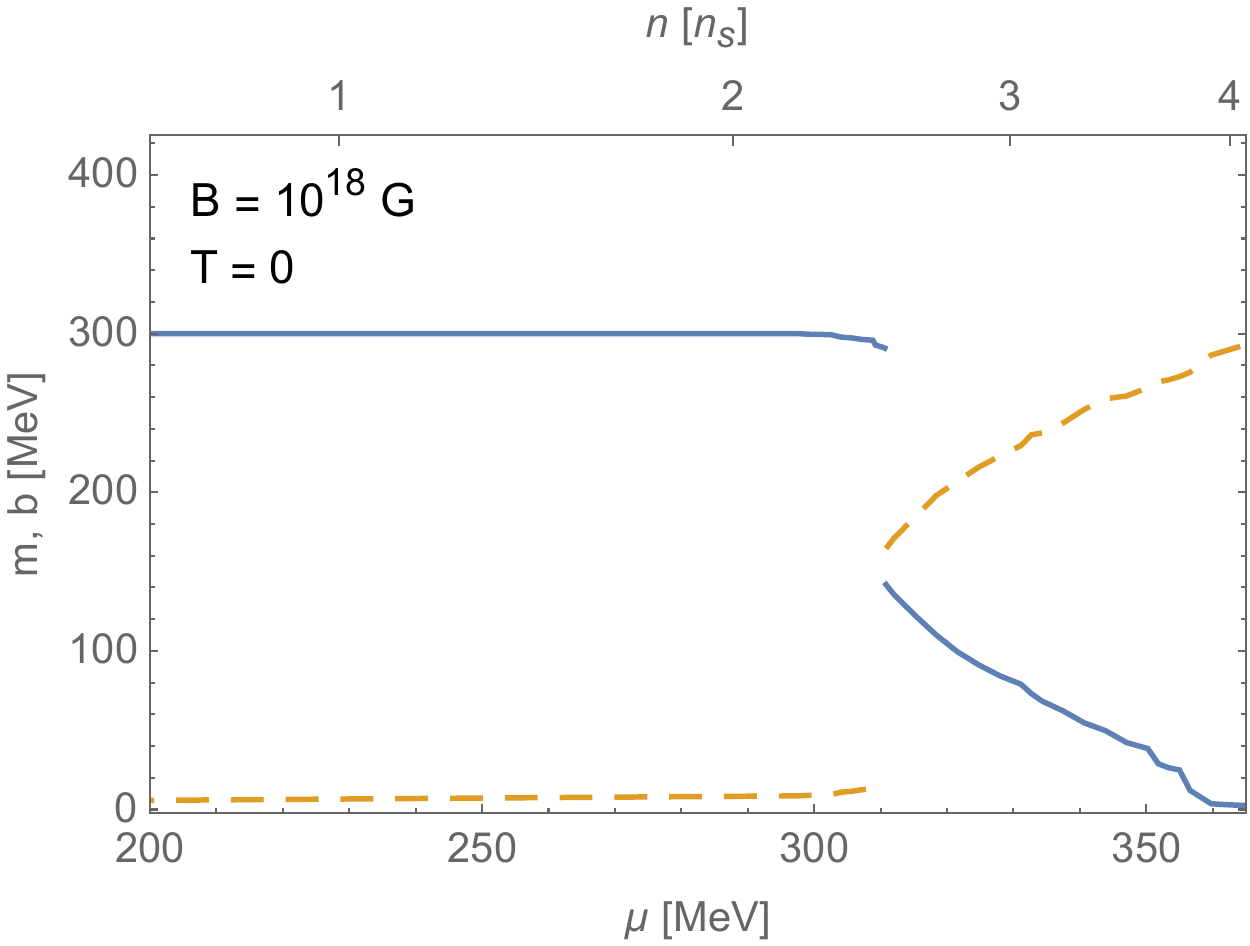}\hfill}
    \qquad
    \subfloat{
    \includegraphics[width=.42\textwidth]{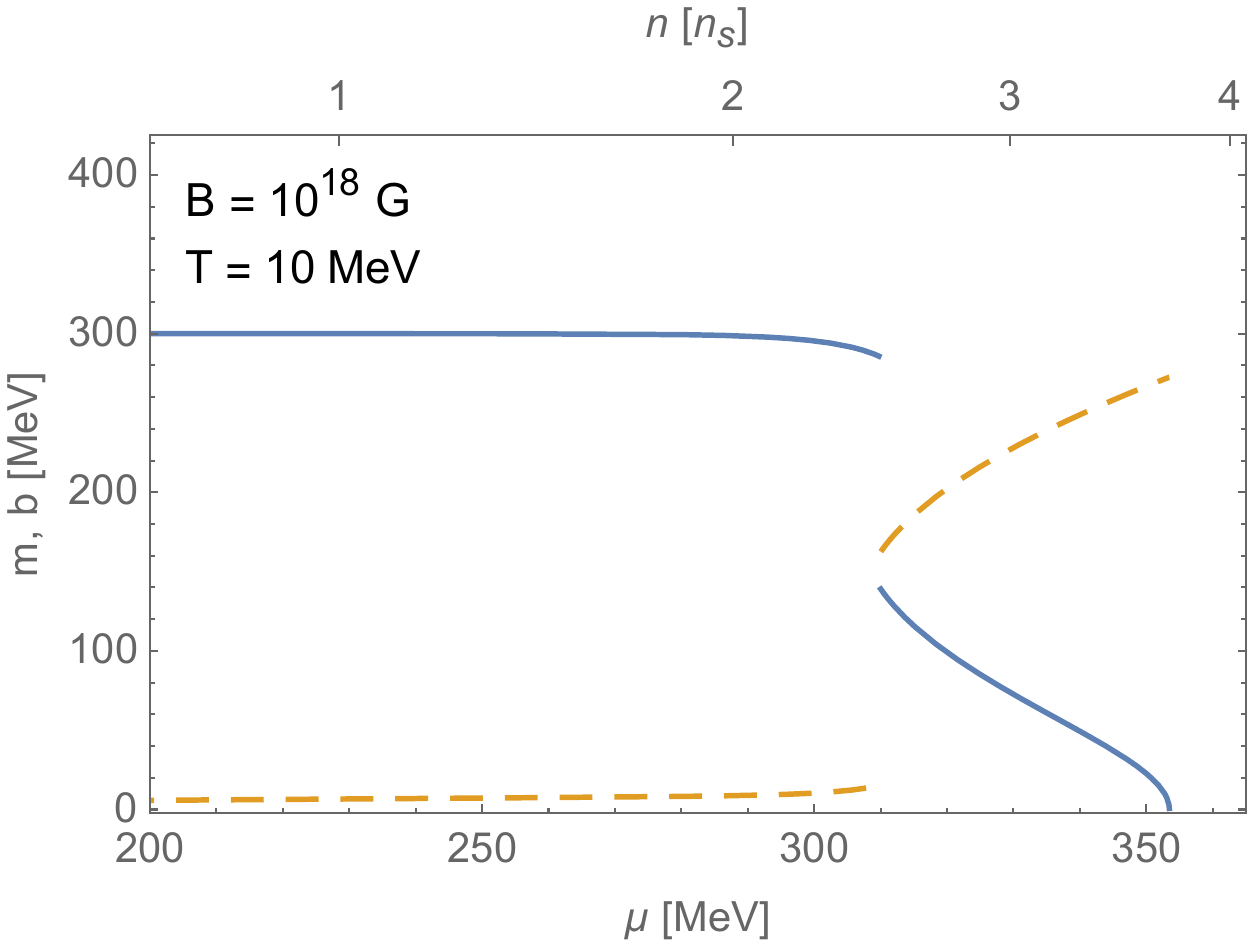}\hfill}\\
    \subfloat{
    \includegraphics[width=.42\textwidth]{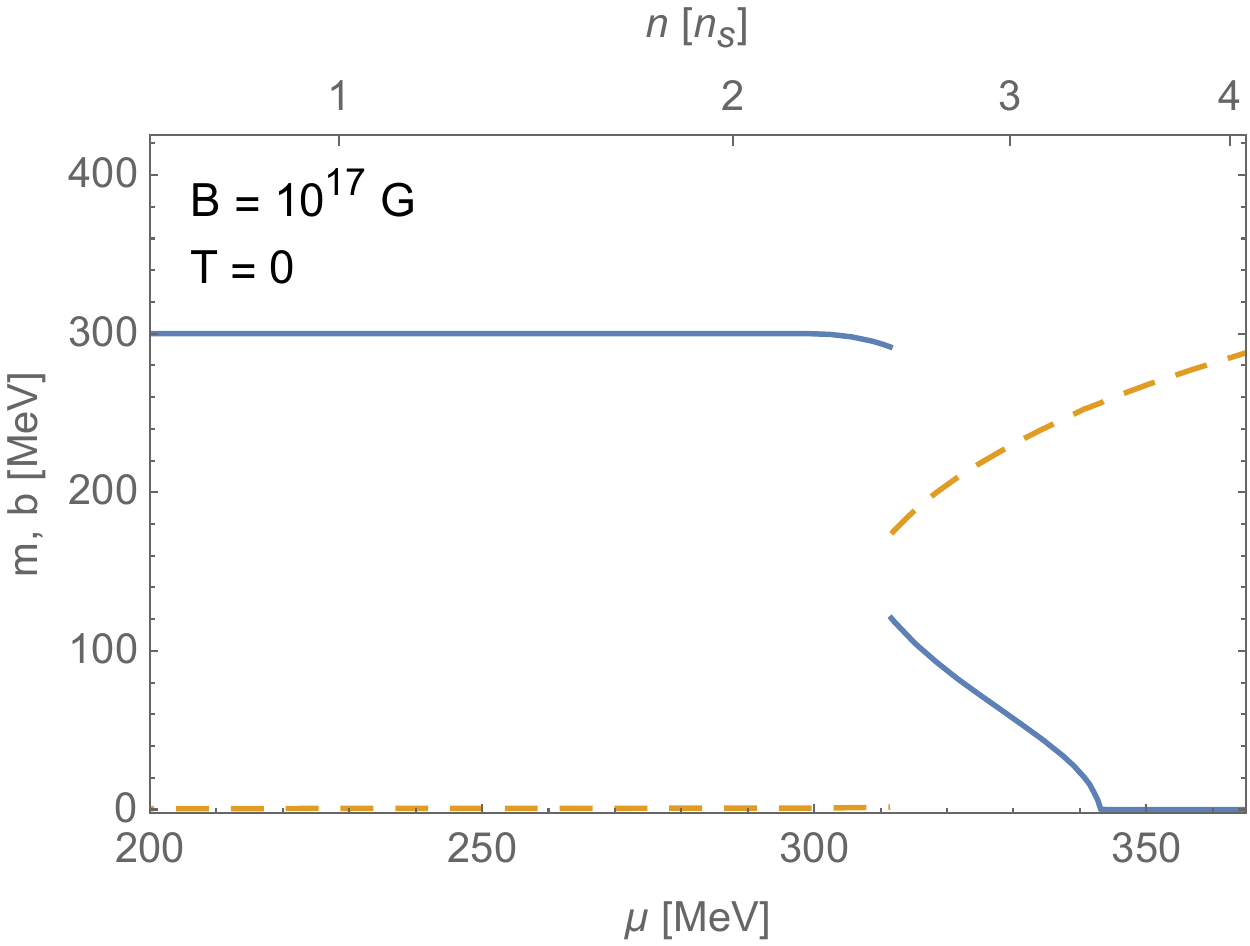}\hfill}
    \qquad
    \subfloat{
    \includegraphics[width=.42\textwidth]{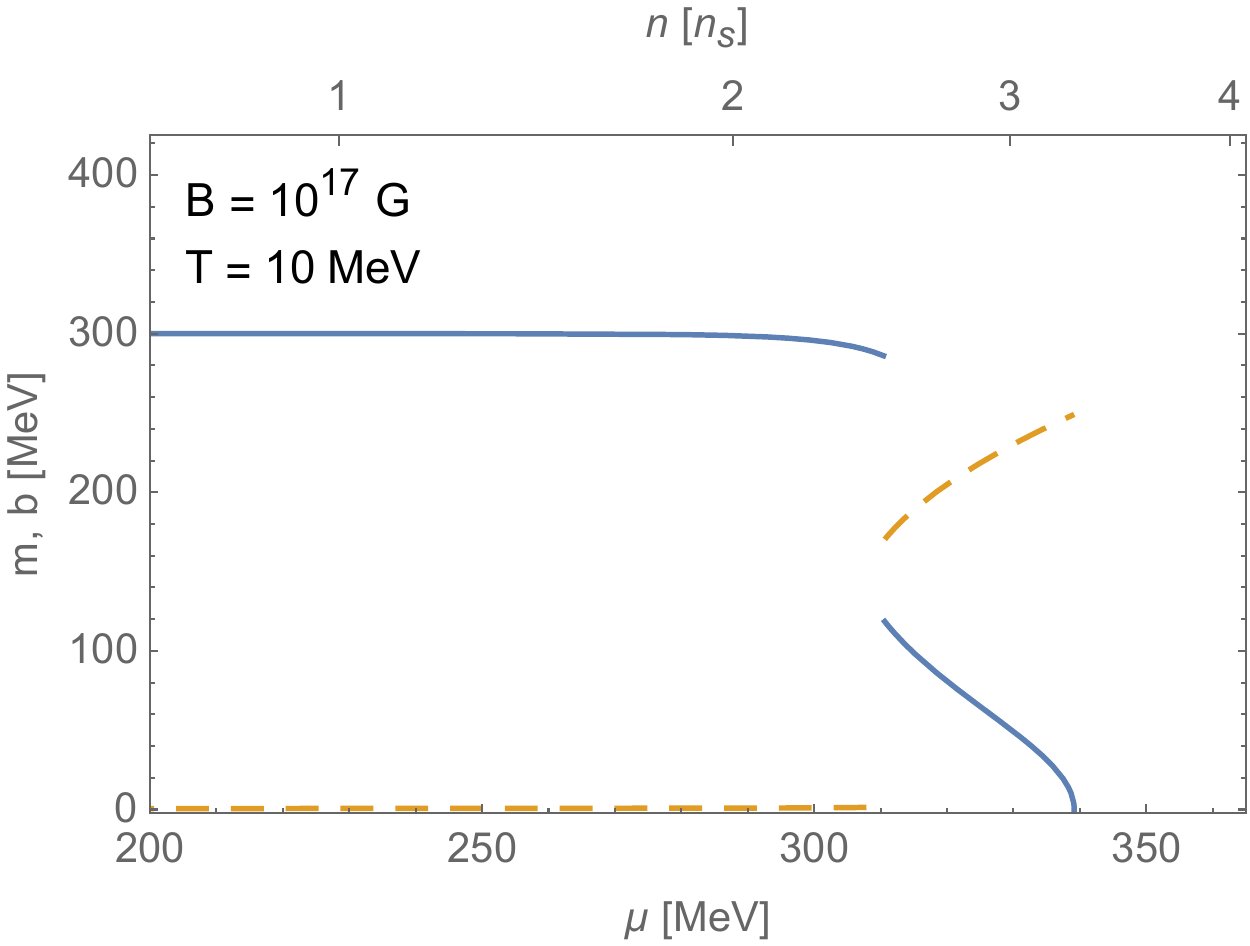}\hfill}
    \caption{Order parameters $m$ (solid) and $b=q/2$ (dashed) plotted against quark chemical potential, with baryonic density in multiples of the saturation density $n_s$ indicated on the top axis. Data are computed by minimizing the exact free energy given by Eqs. (4)--(11) of \cite{MDCDW-3}. Panels show data at $10^{18}$ G (upper) and $10^{17}$ G (lower), as well as zero temperature (left) and $T=10$ MeV (right).}
    \label{mbCurves}
\end{figure}
Fig.\ref{GL6to30} shows the behavior of the threshold temperature with the quark chemical potential (bottom axis) and the corresponding baryon density (top axis) for magnetic field $B=10^{18}$ G. Each colored line indicates the $T_\text{thr}$ found at a given order of the GL expansion. We can see that the sixth order is not very accurate. With increasing order, the results begin to converge, becoming very accurate for $N\geq 20$. After that, the results are indistinguishable in the entire density region and thus are physically reliable, demonstrating the convenience of the large-$N$ GL expansion developed here. Accordingly, for the rest of the calculations presented in this section, we will use a $30$th-order GL expansion. This is also consistent with previous studies \cite{MDCDW-3} that verified the accuracy of the 20th and higher-orders GL expansion compared to exact numerical calculations. 

As shown in \cite{KlimenkoPRD82, PLB743, Topological-Transport-1, Topological-Transport-2}, the MDCDW phase exists for a large range of chemical potentials, including the low chemical potential region, where a Fermi surface has not yet been formed. At low $\mu$, the chiral condensate is almost the same as in vacuum, except for a small modulation (long wavelength) generated by an anomalous term in the thermodynamic potential that comes from the asymmetric spectrum. Such behavior is displayed in Fig. \ref{mbCurves} which shows the condensate's magnitude $m$ and modulation $b=q/2$ vs. $\mu$ at $T=0$ and $T=10$ MeV for different magnetic field values. These results were obtained using the exact thermodynamic potential of the MDCDW phase explicitly given in \cite{Topological-Transport-2, MDCDW-3}. In all the cases, the inhomogeneity becomes noticeable at $\mu\geqslant320$ MeV, or equivalently at $n\geqslant2.5 n_s$. We highlight the dynamical parameters' insensitivity to temperature variations within a range that includes and amply exceeds the characteristic temperatures of old NSs ($\sim 8.6$ keV). 

Another interesting feature of these curves is the presence at $T=0$ of what has been called a remnant mass \cite{MDCDW-3}. The remnant mass indicates the persistence of the condensate, albeit with quite a small magnitude, in regions of relatively high densities (bottom right corners of left panels in Fig. \ref{mbCurves}). This is the only feature not captured by an accurate GL expansion, as discussed in \cite{MDCDW-3}. The profile of $m$ calculated using a reliable ($N \geq 20$) GL expansion for the parameters in Fig. \ref{mbCurves} fully overlaps with that of this figure, except that in the GL case, $m$ vanishes near 360 MeV in the upper left panel and at about 342 MeV in the lower left panel, just at the points where the remnant mass starts.

\begin{figure}
\begin{center}
    \subfloat{
    \includegraphics[width=.42\textwidth]{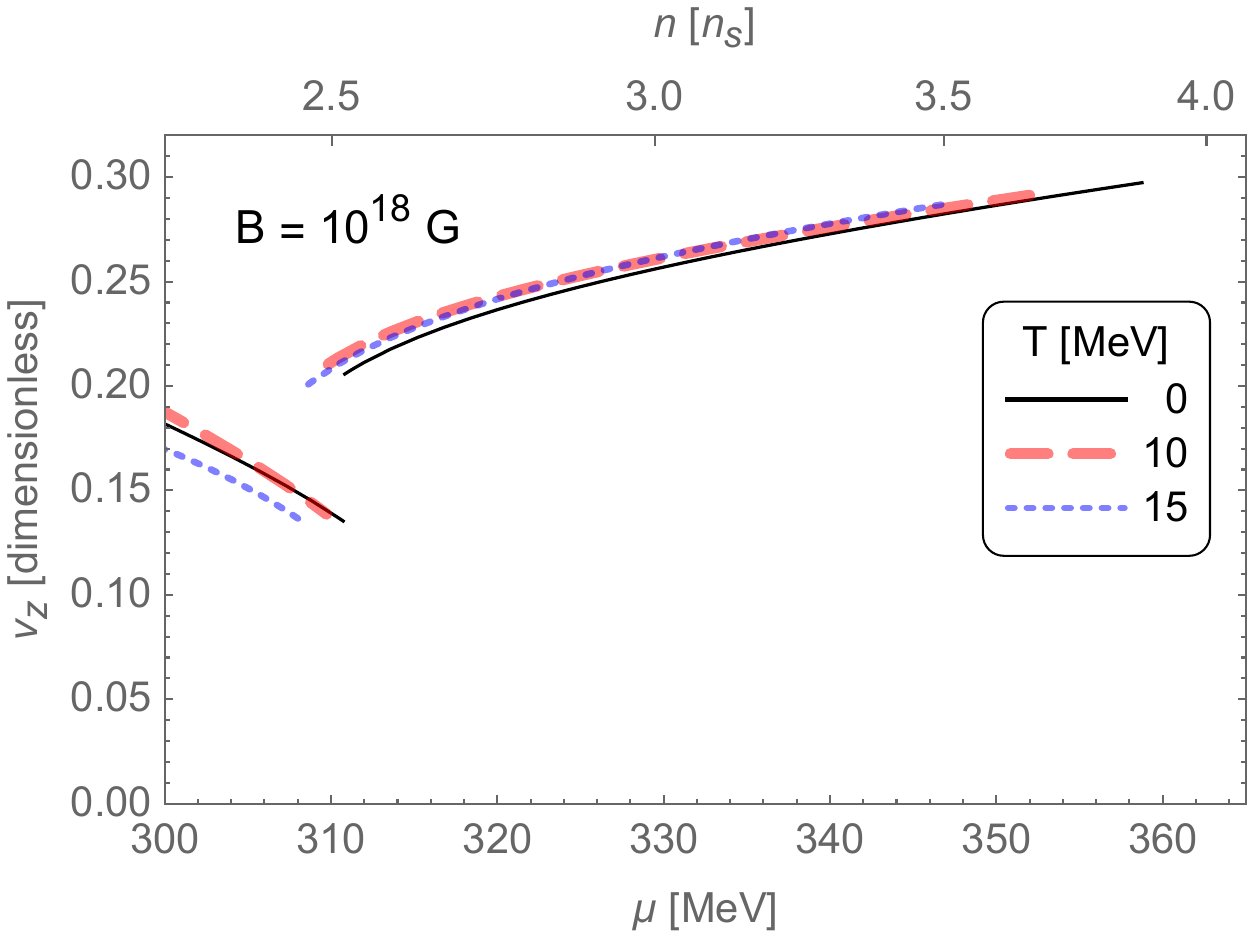}\hfill}
    \qquad
    \subfloat{
    \includegraphics[width=.42\textwidth]{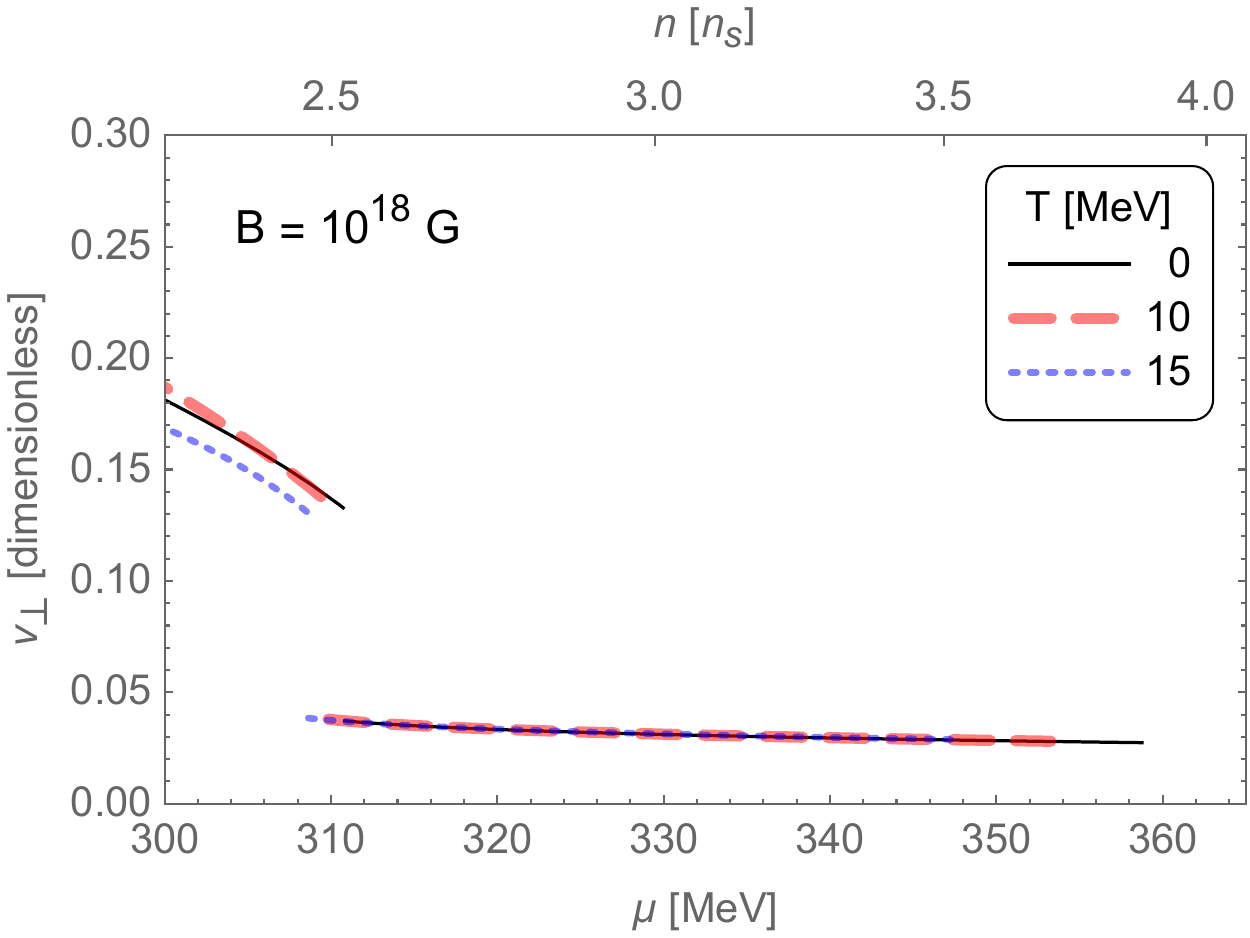}\hfill}\\
    \subfloat{
    \includegraphics[width=.42\textwidth]{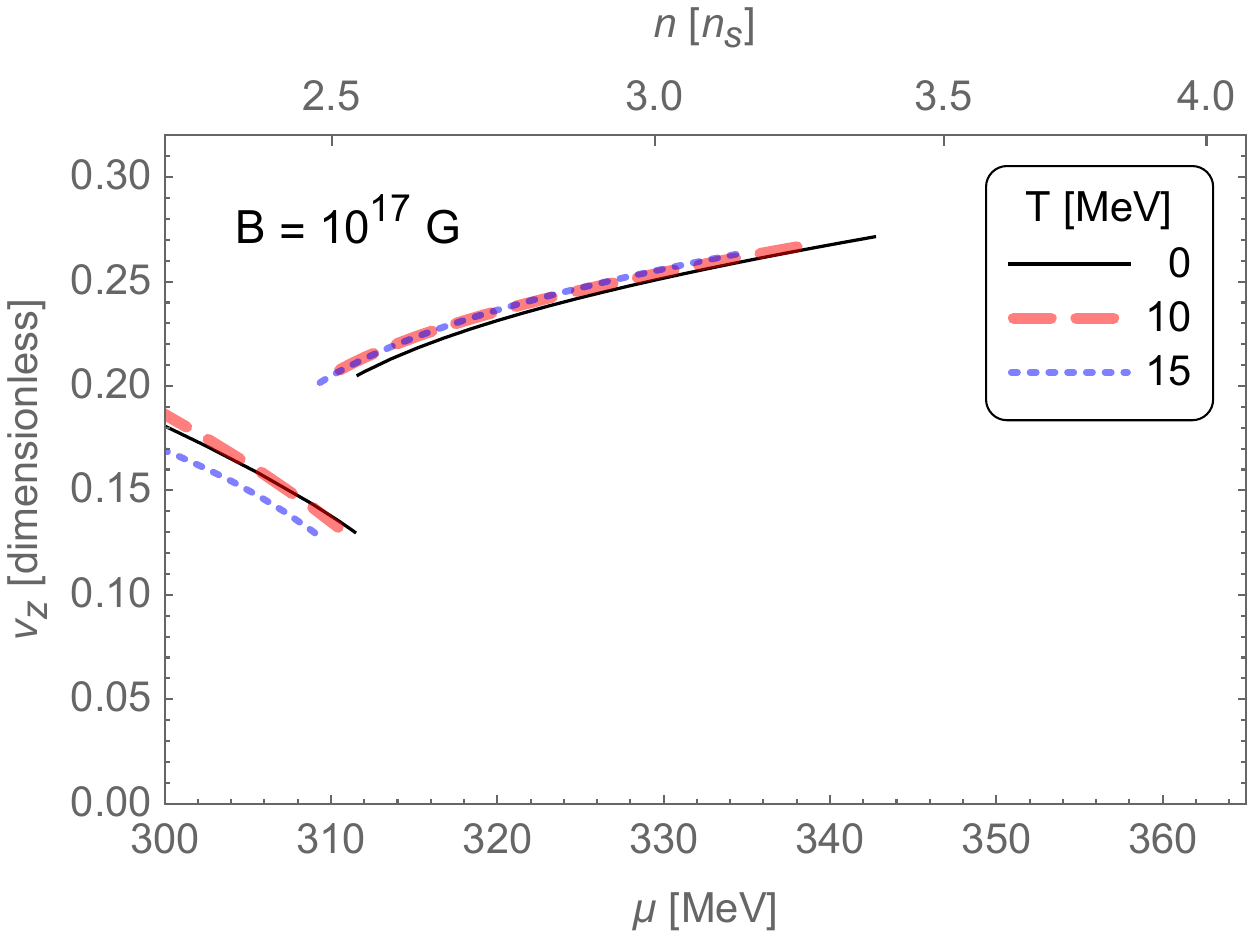}\hfill}
    \qquad
    \subfloat{
    \includegraphics[width=.42\textwidth]{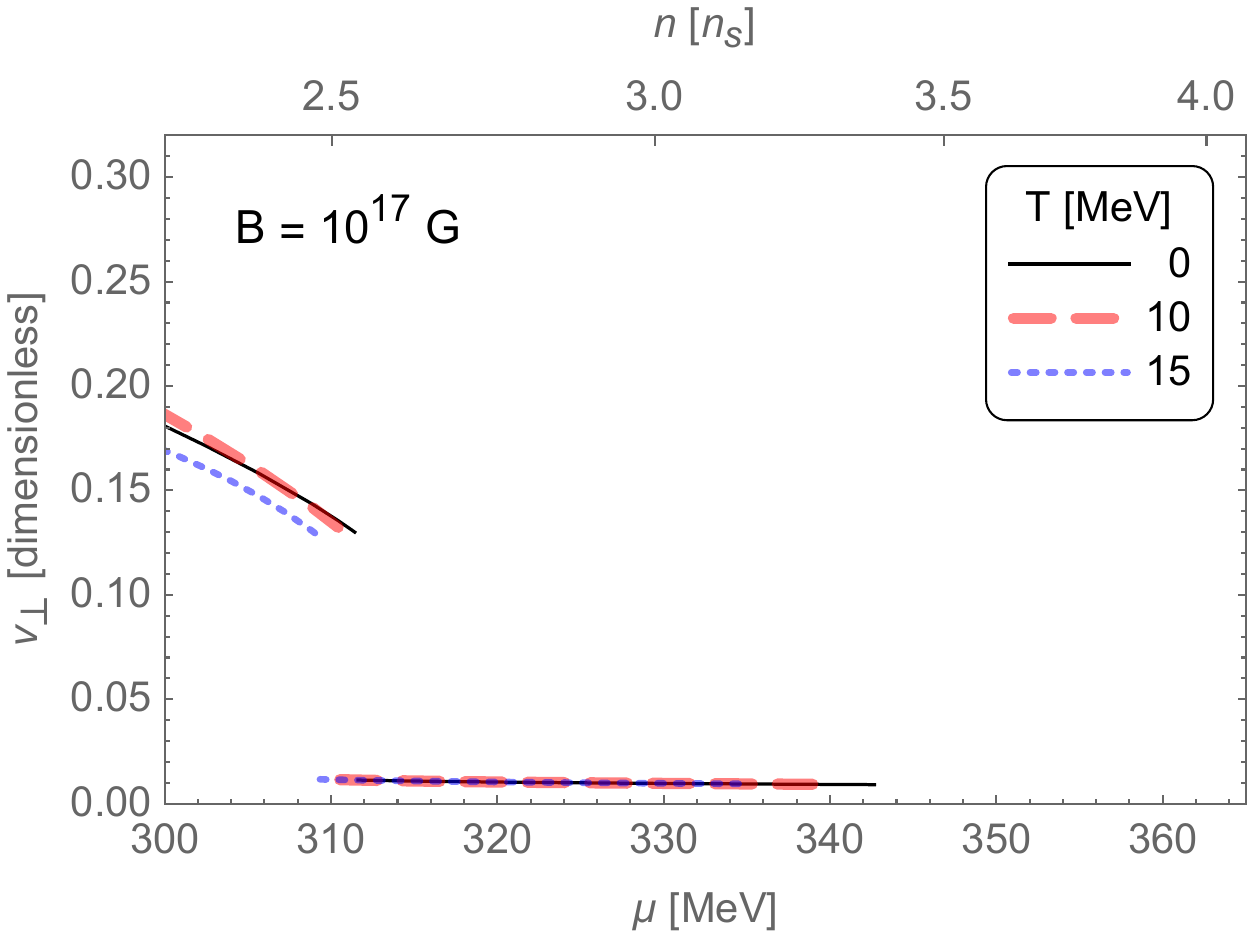}\hfill}
   \caption{Parallel ($v_z$) and transverse ($v_\bot$) group velocities vs. quark chemical potential (bottom axis) and corresponding baryonic density in multiples of the saturation density $n_s$ (top axis) for different magnetic field and temperature values.}
 \label{velocities}
 \end{center}
  \end{figure}

In Fig.\ref{velocities}, we plotted the parallel $v_z$ and transverse $v_\bot$ group velocities vs. chemical potential at various temperatures and magnetic fields. The first thing to notice is a jump in both of them at densities that correlate with the jump in the dynamical parameters (Fig. \ref{mbCurves}). The second observation is that in the low $\mu$ region, the velocities display the same magnetic-field independence shown by the condensate magnitude $m$ \cite{MDCDW-3}. Such a behavior, coupled with the very small inhomogeneity at low $\mu$, leads to a lack of anisotropy in the low $\mu$ region,  as can be gathered from the fact that the parallel and transverse velocities are practically equal there. All this is easy to understand from a physical point of view by recalling that at low density, the condensate is basically the same as in vacuum (except for a tiny, anomalous, field-induced inhomogeneity). Magnetic fields as large as a few times $10^{18}$ G are not big enough to significantly affect the condensate in this region. The low-density region is therefore driven by the vacuum pairing (quark-antiquark pairing), and that is reflected in all the physical parameters, including the condensate, group velocities, and also, as will be seen below, in $T_c$ and $T_\text{thr}$. 

With increasing density, there is a point where the velocities jump so that $v_z$ increases and $v_\bot$ decreases, displaying a sharp anisotropy. After that, each changes smoothly with increasing chemical potential so that $|v_zv_\bot|$ remains roughly constant. Just as expected, the curves stop at the chemical potential where the condensate vanishes using the GL expansion. As for $B$, $v_z$ hardly changes with $B$, while $v_\bot$ decreases with decreasing magnetic field, consistent with the fact that at $B=0$, $v_\bot=0$ and the system exhibits the LP instability.

From the point of view of the fluctuations, the behavior of the velocities can also be understood in terms of the mixed character of the fluctuations and the fact that in the low-density region where the condensate is essentially a vacuum condensate with a tiny modulation, the NG modes are almost pionic, so no much anisotropy should be expected in the group velocities. On the other hand, in the higher density region, after the jump in the dynamical parameters, the particle-hole pairing becomes significant, leading to a sizable condensate modulation and to NG bosons that are phonon-like, all properties that manifest in the anisotropy of the velocities. 

\begin{figure}[h]
    \centering
\includegraphics[width=.43\textwidth]{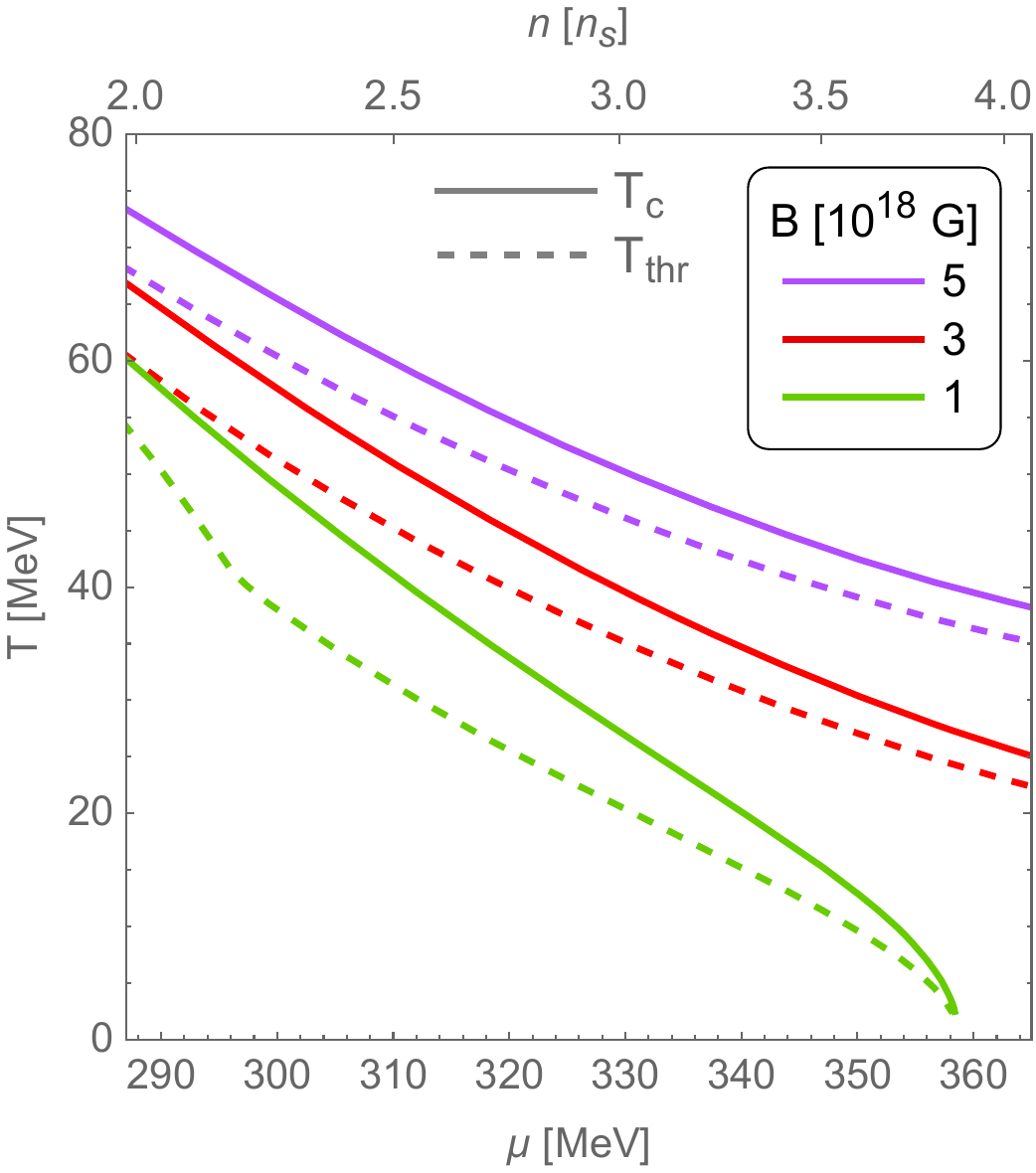}\hfill
    \caption{Comparison of critical and threshold temperatures at three large magnetic field values.}
    \label{TthrLargeB}
\end{figure}

\begin{figure}[h]
    \centering
\includegraphics[width=.24\textwidth]{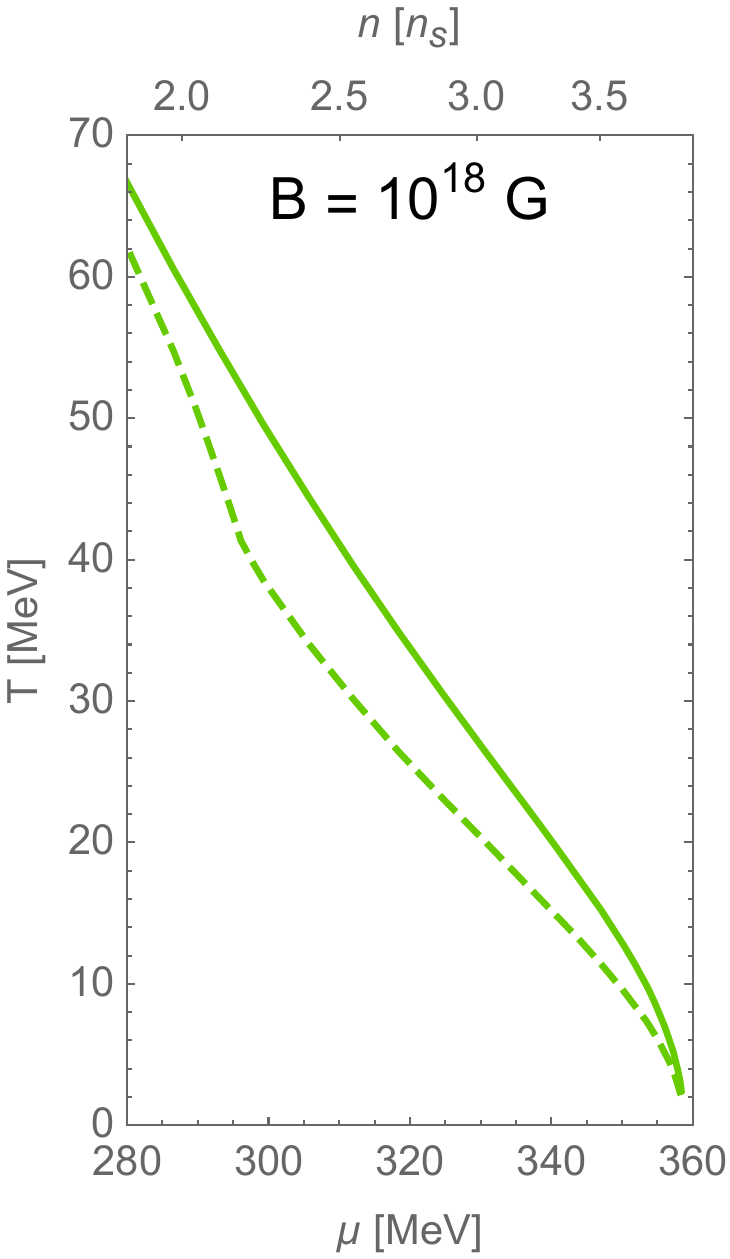}
\includegraphics[width=.24\textwidth]{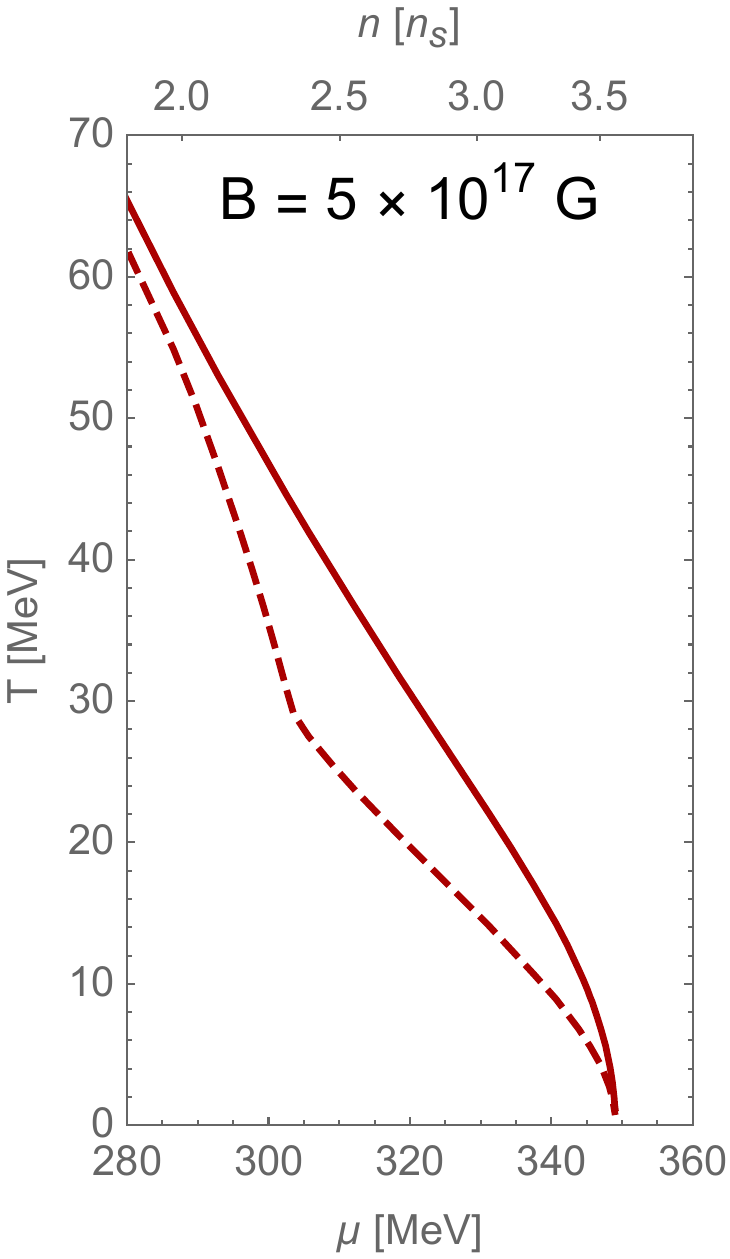}
\includegraphics[width=.24\textwidth]{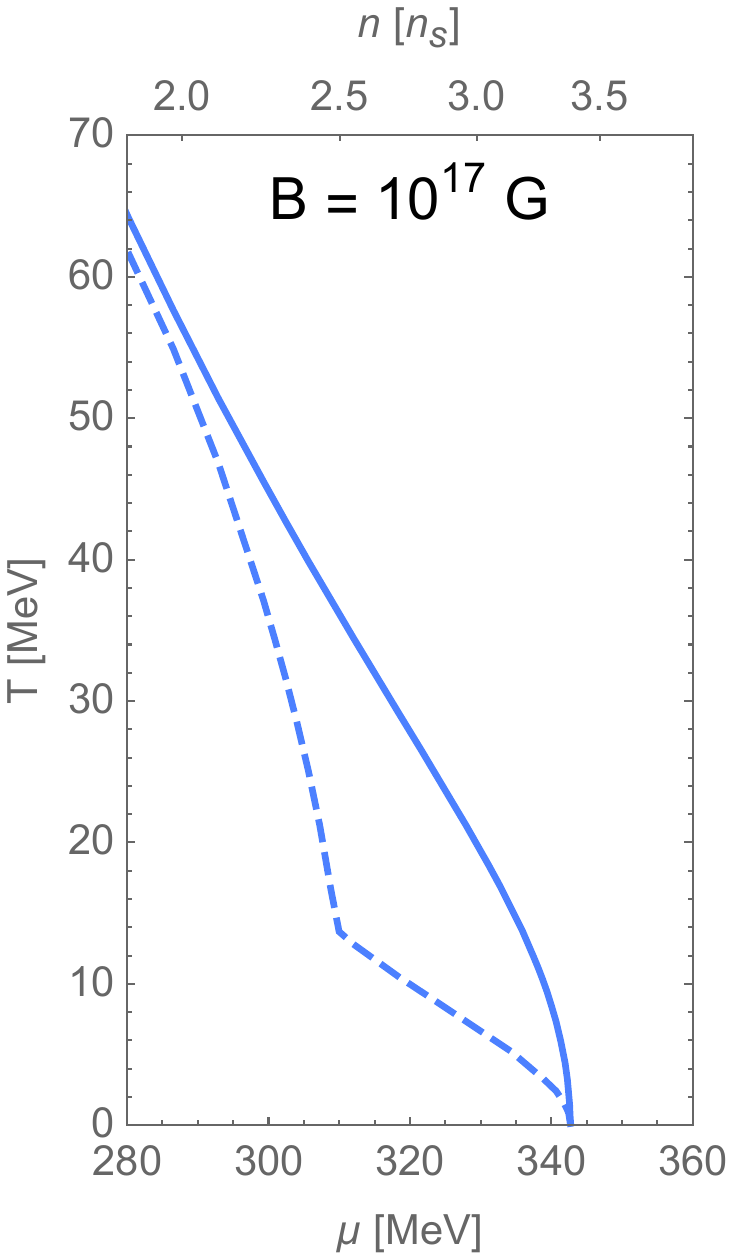}
\includegraphics[width=.24\textwidth]{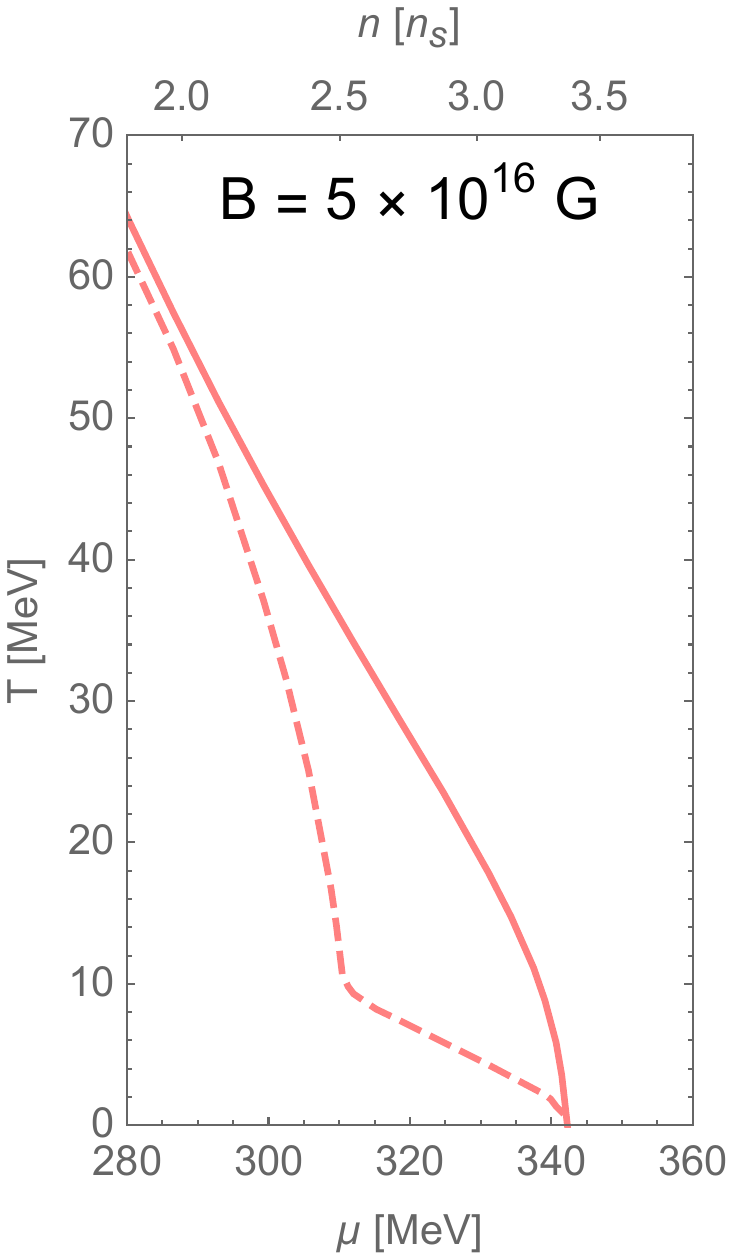}
\hfill
    \caption{Comparison of critical and threshold temperatures at magnetic fields ranging from $5\times10^{16}$--$10^{18}$ G. Solid (dashed) curves show the critical (threshold) temperature. As the magnetic field strength decreases, the threshold temperature decreases---both absolutely and relative to $T_c$---over the range $\mu\gtrsim 312$ MeV, reflecting the more significant effects of the fluctuations at smaller magnetic fields.}
    \label{TthrSmallB}
\end{figure}

\begin{figure}[h]
    \centering
\includegraphics[width=.40\textwidth]{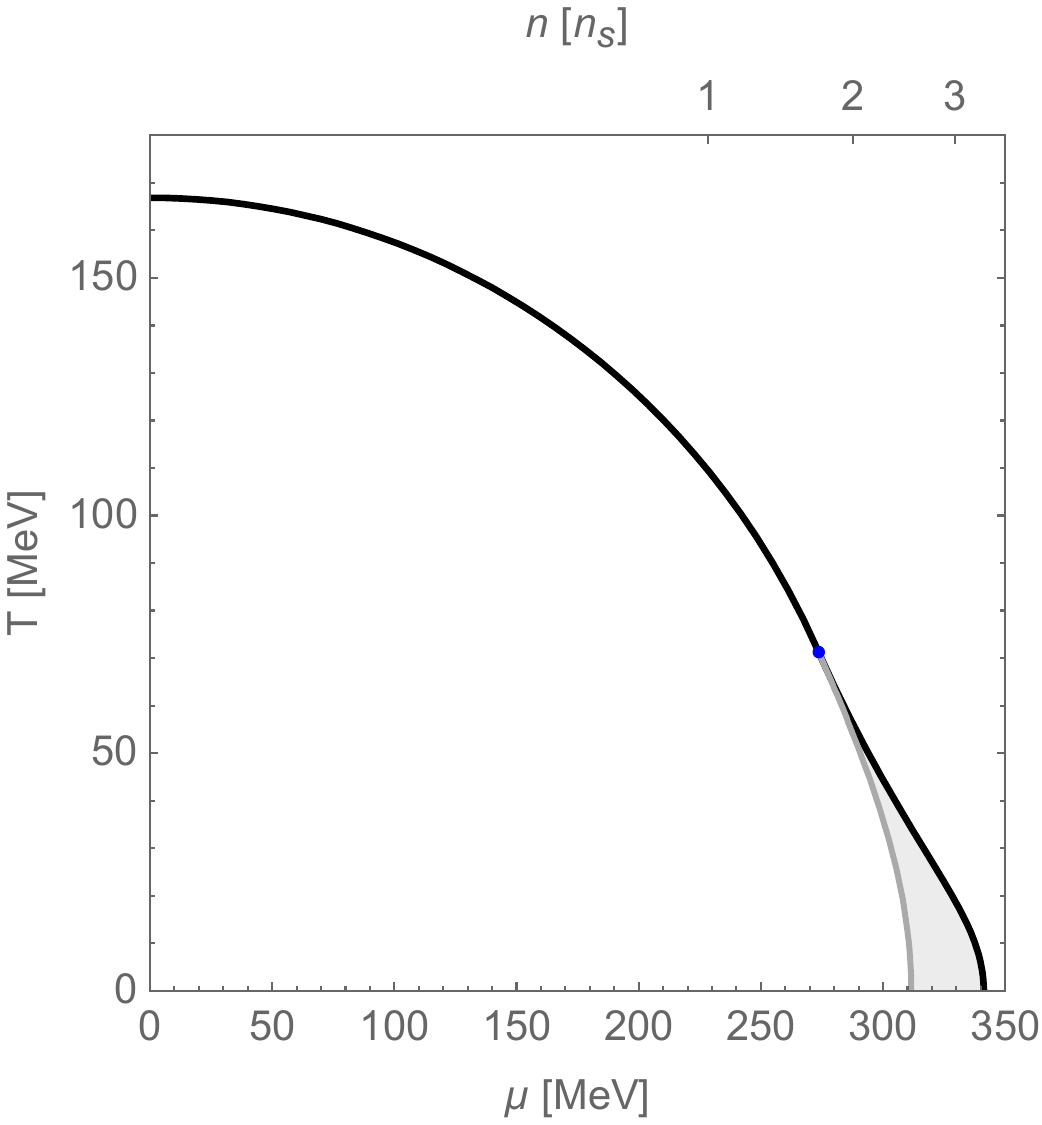}
\includegraphics[width=.40\textwidth]{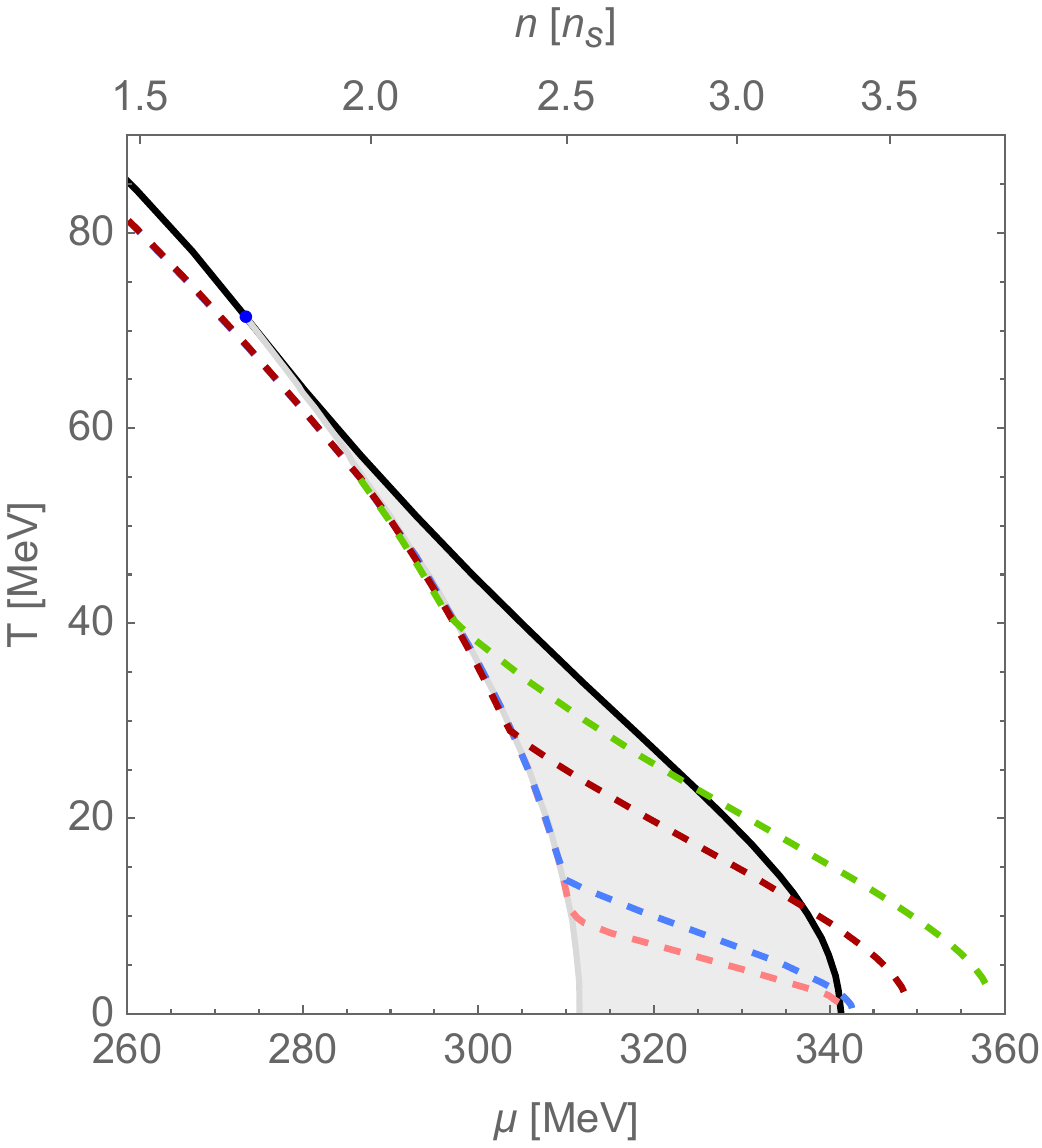}
\hfill
    \caption{Left: Critical temperature (black) at $B=0$. The shaded region indicates where the condensate is inhomogeneous, and the gray line demarcates a first-order transition. Right: Zoomed-in view of the inhomogeneous region at zero $B$ overlaid with threshold temperature curves at finite $B$ for $10^{18}$, $5\times10^{17}$, $10^{17}$, and $5\times10^{16}$ G (top to bottom).}
    \label{InhomRegion}
\end{figure}

\begin{figure}
    \centering
    \includegraphics[width=.43\textwidth]{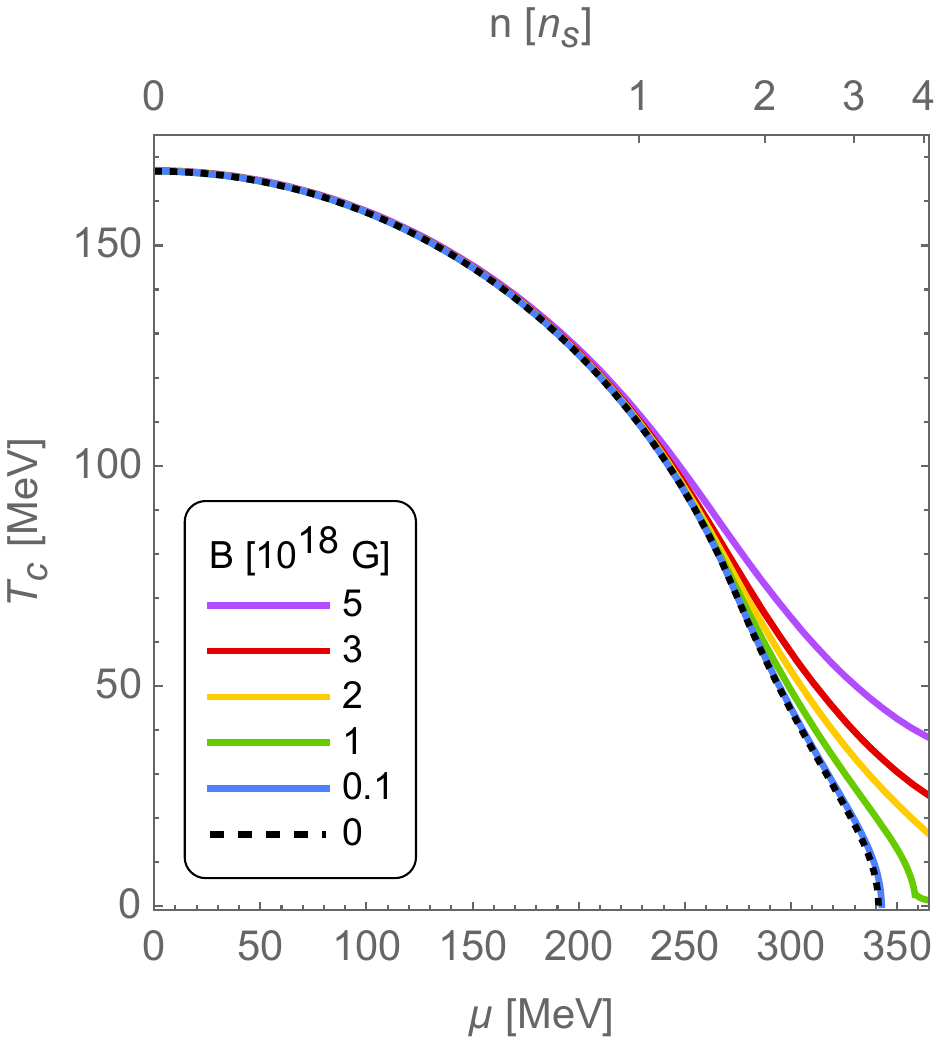}
    \caption{Critical temperature curves plotted against chemical potential at various magnetic field strengths.}
    \label{TcCurves}
\end{figure}

In Fig. \ref{TthrLargeB}, we plot the threshold temperature $T_\text{thr}$, at which the fluctuations become large enough to alter the long-range order, and the critical temperature $T_c$, at which the chiral symmetry is restored; these quantities are both plotted against the quark chemical potential for several multiples of $B=10^{18}$ G. Fig. \ref{TthrSmallB} is similar to Fig. \ref{TthrLargeB}, only at smaller field values that range from $5\times 10^{16}$--$10^{18}$ G in separate panels. We observe that in all these cases $T_\text{thr} < T_c$, as it should be, since the threshold temperature only makes sense in the region where the modulated condensate exists. The two temperatures only overlap when $m=0$ since once the condensate ceases to exist the phonon fluctuations cease to exist too. 

In all cases, a kink in $T_\text{thr}$ moves to the left with increasing magnetic field strength. The kink position correlates with the first-order transition shown in  Fig. \ref{mbCurves} by the jump of the dynamical parameters at some chemical potential. In contrast, the plots of the critical temperature versus $\mu$ have no kink because the transition to restore the chiral symmetry is always second-order. Another interesting observation is that there is a common path traced by all $T_\text{thr}$ curves at smaller $\mu$ before they hit the kink from the left and then branch off to the right of the kink, where $T_\text{thr}$ decreases at a different rate with increasing $\mu$, as expected from the behavior of $|v_zv_\bot |$ and $m$. 

In Fig. \ref{InhomRegion}, in the left panel, the gray-shaded region in the temperature-density plane shows where the DCDW condensate exists at $B=0$. The gray line demarcates a first-order phase transition from the homogeneous condensate to the inhomogeneous one. The dark solid line is $T_c$, which separates the chirally broken, inhomogeneous phase from the chirally symmetric phase. In the right panel, we zoom in on the previous region and superimpose the $T_\text{thr}$ curves calculated at magnetic fields in the interval $5\times10^{16}$--$10^{18}$ G, with colors matching those of Fig. \ref{TthrSmallB}. As mentioned above, as the chemical potential decreases to the left of the kink, all the $T_\text{thr}$ curves approach a single line indicating the magnetic-field independence of this parameter at low densities. A similar thing happens with the critical temperature which becomes field-independent in the low-density region, approching its zero-field value, as shown in Fig. \ref{TcCurves}. Notice the tendency in Fig. \ref{TthrSmallB} of $T_\text{thr}$ to get closer to $T_c$ as $\mu$ decreases. All this correlates with a condensate mostly driven by vacuum pairing in the low-density region, and fluctuations that are essentially pionic there. This pattern is also consistent with the lack of anisotropy in the group velocities and the absence of phonons in the low-density region at $B=0$. On the other hand, the profiles of the $T_\text{thr}$ curves to the right of the kink indicate that the phonon fluctuations become sizable at larger $\mu$. We underline that despite the effect of the phonon fluctuations at larger densities, there is a significant range of densities commensurate with NS densities where the long-range order is robust against thermal fluctuations for magnetic fields $\sim$ $10^{16}$-$10^{18}$ G, which are reasonable field strengths for the interior of magnetars. 

It is worth noting the connection between the LP instability and the Mermin-Wagner-Hohenberg-Coleman Theorem (MWHCT) \cite{Coleman Theo}. According to the MWHCT, spontaneous breaking of a global continuous symmetry cannot occur at $d \leqslant 2$, whether in QFT or at finite temperature, because it would lead to infrared divergent correlation functions of the Goldstone bosons or, equivalently, to divergently large fluctuations.

Meanwhile, the LP instability states that single-modulated condensates in 3+1 dimensions are unstable against thermal fluctuations, regardless of the temperature. This is due to having a soft mode in the transverse direction in these systems. Such a soft mode exhibits $v_\bot= 0$ and produces an extreme anisotropy in the propagation of the fluctuations. Therefore, even though the original fermion theory is 3+1-dimensional, the spectrum of the phonon fluctuations is essentially 1+1-dimensional. This fact implies that the momentum integration in the expression of the average square fluctuation exhibits the same infrared behavior that characterizes 1+1-dimensional theories of fermions subject to the MWHCT.

It is remarkable that if the ground state breaks $T$, a magnetic field can alter the fluctuation spectrum in such a way that single-modulated phases in 3+1 dimensions can be free of LP instability. This results in long-range orders that remain robust against thermal fluctuations up to very high temperatures. In such cases, the analogy with the MWHCT ceases since the fluctuations now propagate in 3+1 dimensions, as $v_\bot \neq 0$, and the integrals in momentum become infrared finite. This is precisely the situation with MDCDW. Here, the coexistence of a $T$-odd ground state and a magnetic field allows the emergence of $b$-terms in the GL expansion, ensuring the lack of LP instability. As mentioned above, these terms have a topological origin. We know for a fact that they only get contributions from the LLL, whose spectrum is asymmetric and hence topological \cite{Niemi}. The asymmetric spectrum is also responsible for an anomalous baryon number proportional to the Atiyah-Singer index \cite{PLB743}. We believe that there is a deeper physical connection among all these features, which goes beyond the MDCDW case. We expect that when a magnetic field is present, at least for theories with the symmetries of Lagrangian (\ref{L_NJL_QED}), any of the following features implies the others and prevent the LP instability: \textit{i}) a $T$-breaking ground state, \textit{ii}) an anomalous baryon number, or \textit{iii}) an asymmetric fermion spectrum. 

We argue that, in general, a magnetic field alone is not enough to cure the LP instability in a single-modulated phase, despite some claims in the literature to the contrary \cite{HidakaPRD92}. To understand this, let us first recall that at $B=0$, the free energy of the fluctuations cannot contain terms proportional to $(\partial u/ \partial x)^2$ or $(\partial u/ \partial y)^2$, as these terms would come from rotations about $y$ or $x$ respectively, but those rotations cannot cost energy due to the system's isotropy \cite{Baym}. Therefore, the argument that $v_\bot= 0$ at $B=0$ is a manifestation of the isotropy. However, the lack of isotropy is a necessary but not sufficient condition for $v_\bot\neq 0$. For instance, in the presence of a magnetic field, a term of the form $\alpha B_z\partial_z u$ in the fluctuation's free energy could in principle lead to $v_\bot\neq 0$ \cite{HidakaPRD92}, but such a term is forbidden if $\alpha$ is $T$-even. Thus, $\alpha$ must be $T$-odd for such a term to be allowed, but $\alpha$ can only be odd under $T$ if the ground state breaks that symmetry. Hence, the argument of \cite{HidakaPRD92} is only valid for theories where the condensate breaks $T$ symmetry. Additionally, since any term linear in $B$ can only come from the LLL, we see once again how the two features already underlined as relevant, dimensional reduction and $T$-breaking, also emerge within this more general argument.

Our results have demonstrated that magnetars' cores could be in the MDCDW phase, given that for the characteristic fields, temperatures, and densities of those stars, MDCDW is realizable in principle, and the thermal fluctuations would not be big enough to wash out the long-range order. This strengthens the case for this phase as a feasible candidate for the superdense matter of magnetized compact objects.

\section{Concluding Remarks}
\label{secIV}
 
In this paper, we investigated the stability of the MDCDW phase against thermal phonon fluctuations. We generalized the method developed in \cite{MDCDW-2} to the case of an $N$th-order GL expansion and then used the formulas for the $N$th-order GL coefficients derived in \cite{MDCDW-3} to compute the threshold temperature to high accuracy. Our results show that for magnetic fields of order $10^{18}$ G, the inhomogeneous condensate of the MDCDW phase remains stable against fluctuations over most of the parameter space in which it is energetically favored. For fields of order $10^{17}$ G, the condensate remains stable up to a significant fraction of the critical temperature, leaving some regions of only quasi-long-range order before chiral restoration occurs. 

Let us highlight that, as far as compact star applications are concerned, the temperature scales involved here are several orders of magnitude greater than those of old NSs, which are generally on the order of keV rather than MeV. Therefore, the conditions of a typical old NS appear favorable to MDCDW at densities in the range of about 2--3.5 $n_s$, at least in magnetic fields of order $10^{16}$ G. In larger magnetic fields, this range of densities widens, and for fields $B\gtrsim2\times10^{18}$ G, MDCDW is stable and preferred over the chirally symmetric state over the entire range of densities 2--11 $n_s$. On the other hand, temperatures can be much higher in the short-lived remnants of NS mergers, reaching several tens of MeV, but more extreme magnetic fields and densities also accompany these events. Recent studies exploring a large ensemble of model-independent equations of state that incorporate data from multimessenger observations and the Neutron Star Interior Composition Explorer (NICER) \cite{Annala1,Annala2} have found compatibility with the presence of quark matter cores in massive NSs. Therefore, the realization of MDCDW in the remnant products of these violent collisions remains plausible.

The study carried out in this paper fits into a natural progression of investigations into the MDCDW phase, where we continue to check for characteristics that permit its feasibility in real-world scenarios. Previous work focusing on the mean-field theory has found this phase to be resilient in several senses: It remains the ground state up to a significant critical temperature at extremely low and high densities, and it retains a nonvanishing remnant mass and a large modulation parameter at higher intermediate densities \cite{MDCDW-3}. Other studies have established compatibility between the predicted physical properties of this phase, such as heat capacity \cite{PRD20} and maximum stellar mass \cite{InhStars}, with observational constraints.

Here, we have shown that this resilience persists beyond mean-field theory and carries over to the phase's stability against thermal fluctuations. A natural follow-up task is to compare the free energy of this phase with those of the 2SC and mixed phases to determine which one is energetically preferred when a magnetic field is present. Future theoretical work in this direction, combined with further constraints from multimessenger observations and numerical simulations, will continue to shed light on the relevance of inhomogeneous condensates to astrophysics. Thus far, the MDCDW phase continues to emerge as a robust candidate for the microphysical description of matter in the intense environment of compact stars.

\acknowledgments
This work was supported in part by NSF grant PHY-2013222. 
 
\appendix

\section{Derivation of the formulas containing $\bm{\nabla^kM}$}
\label{app:d^k M}

In this Appendix, we derive the formulas given in (\ref{d^kM2}) and (\ref{Im d^kM}). Inserting $M_0(z)=me^{iqz}$ into (\ref{perturbation}) gives
\begin{equation}
     M(x)=me^{iqz}\left[1+iqu(x)-\frac12q^2u^2(x)\right],
\end{equation}
and it is straightforward to show that
\begin{equation}
\label{dM}
     \nabla M=iq(\hat z+\nabla u)M,
\end{equation}
where for simplicity we suppressed the spatial parameter $x$. As before, ``$=$'' denotes equality up to second order in $u$, and neglecting terms with more than two overall derivatives. 

We first show that
\begin{equation}
\label{d^k M}
    \nabla^kM=\left[(iq)^k(\hat z+\nabla u)^k+(iq)^{k-1}D^2u\right]M,
\end{equation}
where $D^2u$ denotes a sum of second derivatives of $u$, with real scalar or vector pre-factors according to whether $k$ is even or odd, respectively. The expression $(\hat z+\nabla u)^k$ denotes $[(\hat z+\nabla u)\cdot(\hat z+\nabla u)]^{k/2}$ if $k$ is even, and $[(\hat z+\nabla u)\cdot(\hat z+\nabla u)]^{(k-1)/2}(\hat z+\nabla u)$ if $k$ is odd. (\ref{d^k M}) clearly holds for $k=0$ (interpreting $\nabla^0 M$ as $M$, and noting that the pre-factors in the terms of $D^2u$ can all be zero), and (\ref{dM}) shows that (\ref{d^k M}) also holds for $k=1$. We proceed by induction, treating even and odd cases separately.

Suppose (\ref{d^k M}) holds for $k=2n$, i.e., 
\begin{equation}
\label{eqn:d^2n M}
     \nabla^{2n}M=\left[(iq)^{2n}(\hat z+\nabla u)^{2n}+(iq)^{2n-1}D^2u\right]M.
\end{equation}
Then 
\begingroup
\allowdisplaybreaks
\begin{align}
\label{d^2n+1 M}
     \nabla^{2n+1}M&=\nabla(\nabla^{2n}M)\nonumber\\
     &=\nabla\left[(iq)^{2n}(\hat z+\nabla u)^{2n}+(iq)^{2n-1}D^2u\right]M+\left[(iq)^{2n}(\hat z+\nabla u)^{2n}+(iq)^{2n-1}D^2u\right]\nabla M\nonumber\\
     &=\left[(iq)^{2n}\nabla(\hat z+\nabla u)^{2n}+(iq)^{2n-1}\nabla D^2u
     +(iq)^{2n+1}(\hat z+\nabla u)^{2n+1}+(iq)^{2n}D^2u\right]M,\nonumber\\
\end{align}
\endgroup
where we used (\ref{dM}) in the last step. The term with $\nabla D^2u$ can be neglected because it has more than two overall derivatives. We also have $\nabla(\hat z+\nabla u)^{2n}=\nabla(1+2\pd_zu+|\nabla z|^2)^n=n(1+2\pd_zu+|\nabla z|^2)^{n-1}(2\nabla\pd_zu+\nabla|\nabla u|^2)=2n\nabla\pd_zu$, again neglecting terms with three overall derivatives. Since $\nabla\pd_zu$ contains only second derivatives of $u$, it can be absorbed into $D^2u$, and then (\ref{d^2n+1 M}) reduces to (\ref{d^k M}) for $k=2n+1$, as desired. A similar computation shows that if (\ref{d^k M}) holds for $k=2n+1$, then it holds for $k=2n+2$, completing the proof by induction for (\ref{d^k M}).

Observe that (\ref{d^k M}) contains a sum of pure real and imaginary terms, so the cross terms cancel when calculating $|\nabla^kM|^2$. Moreover, the term involving $|D^2u|^2$ can be neglected. Finally, it is easy to check that $|M|^2=m^2$ up to second order in $u$, and (\ref{d^kM2}) follows immediately from these comments. Eq. (\ref{Im d^kM}) also follows quickly from (\ref{d^k M}). In this case, since $k$ is even, the term involving $D^2u$ turns out to be purely real, and hence it vanishes under the action of Im.


\end{document}